\documentclass[preprint,journal]{vgtc}       % preprint (journal style)

\ifpdf%                                % if we use pdflatex
  \pdfoutput=1\relax                   % create PDFs from pdfLaTeX
  \usepackage{graphicx}                % allow us to embed graphics files
  \DeclareGraphicsExtensions{.pdf,.png,.jpg,.jpeg} % for pdflatex we expect .pdf, .png, or .jpg files
\else%                                 % else we use pure latex
  \usepackage{graphicx}                % allow us to embed graphics files
  \DeclareGraphicsExtensions{.eps}     % for pure latex we expect eps files
\fi%

\graphicspath{{figures/}{pictures/}{images/}{./}} % where to search for the images

\usepackage{microtype}                 % use micro-typography (slightly more compact, better to read)
\PassOptionsToPackage{warn}{textcomp}  % to address font issues with \textrightarrow
\usepackage{textcomp}                  % use better special symbols
\usepackage{mathtools}
\usepackage{times}                     % we use Times as the main font
\usepackage{cite}                      % needed to automatically sort the references
\usepackage{tabu}                      % only used for the table example
\usepackage{booktabs}                  % only used for the table example
\newcolumntype{P}[1]{>{\centering\arraybackslash}p{#1}}
\usepackage[export]{adjustbox}
\usepackage[colorlinks=true,backref=page]{hyperref}
\usepackage{subfig}
\usepackage{graphicx}
\usepackage{rotating}
\usepackage{pifont}
\usepackage{array}

\usepackage[switch]{lineno}

\usepackage[final,authormarkup=none]{changes}
\definechangesauthor[name=akila, color=red]{A-2}
\definechangesauthor[name=alex, color=red]{A1}

\ieeedoi{10.1109/TVCG.2023.3243834}

\onlineid{0}

\vgtccategory{Research}
\vgtcpapertype{application/design study}
\title{RipViz:\\
Finding Rip Currents by Learning Pathline Behavior}

\author{Akila de Silva, Mona Zhao, Donald Stewart, Fahim Hasan Khan, Gregory Dusek, James Davis, and Alex Pang}
\authorfooter{
\item
 Akila de Silva, Mona Zhao, Donald Stewart, Fahim Hasan Khan, James Davis, Alex Pang  is with UC Santa Cruz. E-mail: audesilv@ucsc.edu, yzhao172@ucsc.edu, dolstewa@ucsc.edu, fkhan4@ucsc.edu, davis@cs.ucsc.edu, pang@soe.ucsc.edu
 \item
Gregory Dusek is with NOAA. E-mail: gregory.dusek@noaa.gov.
\item 
Corresponding Authors : Akila de Silva and Alex Pang

}

\shortauthortitle{de Silva \MakeLowercase{\textit{et al.}}: RipViz: Finding Rip Currents by Learning Pathline Behavior}
\abstract{We present a hybrid \added[id=A1]{machine learning and flow analysis} feature detection method, RipViz, \deleted[id=A1]{that combines machine learning and flow analysis} to extract rip currents from stationary videos. Rip currents are \added[id=A1]{dangerous} strong currents that can drag beachgoers out to sea. Most people are either unaware of them or do not know what they look like. In some instances, even trained personnel such as lifeguards have difficulty identifying them. \added[id=A1]{RipViz produces a simple, easy to understand visualization of rip location overlaid on the source video}. With RipViz, we first obtain an unsteady 2D vector field from the stationary video using optical flow. 
\added[id=A1]{Movement at each pixel is analyzed over time. At each seed point,}
\deleted[id=A-2]{short} sequences of \added[id=A-2]{short} pathlines, rather a single long pathline, are \deleted[id=A1]{seeded and} traced across the frames of the video to better capture the quasi-periodic flow behavior of wave activity. Because of the motion on the beach, the surf zone, and the surrounding areas, these pathlines may still appear very cluttered and incomprehensible. Furthermore, lay audiences are not familiar with pathlines and may not know how to interpret them. \deleted[id=A1]{In RipViz,}\added[id=A1]{To address this,} we treat rip currents as a flow anomaly in an otherwise normal \deleted[id=A1]{ocean} flow. To learn about the normal flow behavior, we train an LSTM autoencoder with pathline sequences from normal ocean\added[id=A1]{, foreground, and background movements.} \deleted[id=A1]{flow.
A pathline sequence is simply a series of shorter pathlines seeded from the same point at different times.} During test time, we use the trained LSTM autoencoder to detect anomalous pathlines (i.e., those in the rip zone). The origination points of such anomalous pathlines, over the course of the video, are then presented as points within the rip zone.
\deleted[id=A-2]{In contrast to an earlier method that also uses flow analysis to detect rip currents,} RipViz is fully automated and does not require user input. Feedback from domain expert suggests that RipViz has the potential for wider use.
} % end of abstract

\keywords{Flow visualization, 2D unsteady flow fields, pathlines, LSTM autoencoders, anomaly detection}

\CCScatlist{ % not used in journal version
 \CCScat{K.6.1}{Management of Computing and Information Systems}%
{Project and People Management}{Life Cycle};
 \CCScat{K.7.m}{The Computing Profession}{Miscellaneous}{Ethics}
}

\teaser{
  \centering
  \includegraphics[width=\linewidth]{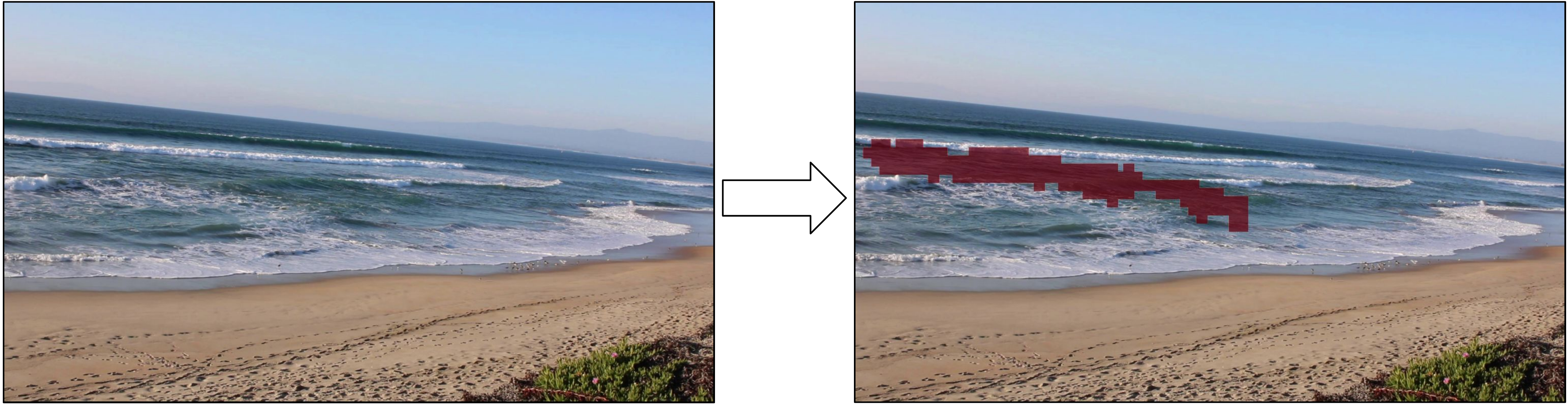}
  \caption{\added[id=A-2]{\textbf{Rip currents are deadly but remain invisible to many:} We propose a feature detection method, RipViz, to make these invisible rip currents visible.} 
  \added[id=A1]{RipViz highlights locations of rip currents as a red region.  This region is determined by finding seed points that produce pathline sequences deviating from normal ocean flow.} \added[id=A-2]{Non-experts and experts alike can use RipViz to visualize rip currents.} \deleted[id=A1]{RipViz analyzes pathline sequences of ocean flow to predict anomalous pathline sequences (i.e., rip currents). The seed points of the anomalous pathline sequences are colored in red.}} 
  \label{fig:teaser}
}

\vgtcinsertpkg

\begin{document}

%% The ``\maketitle'' command must be the first command after the
%% ``\begin{document}'' command. It prepares and prints the title block.

%% the only exception to this rule is the \firstsection command
\firstsection{Introduction}

\maketitle

% \section{Introduction \label{s:intro}} %for journal use above \firstsection{..} instead
Rip currents are powerful, narrow channels of fast-moving water flowing towards the sea from the nearshore \cite{bowen1969rip, castelle2016rip, leatherman11,macmahan11}. 
\added[id=A1]{The speed of seaward rips can be very strong, reaching \added[id=A-2]{two meters} \deleted[id=A-2]{five miles} per second, faster than an Olympic swimmer. They are a dangerous beach hazard that most people do not recognize.  As a result, there are thousands of drownings each year due to rip currents globally \cite{sea_bathing_stats_klein,rip_current_drowning_lushine}.  The goal of this work is to improve public safety by helping the beachgoers {\em see} the rip currents when they are present.}

The mechanism for rip currents is an \deleted[id=A1]{An} increase in the mean water level, referred to as setup, \added[id=A1]{which }occurs when waves break against the shore. This setup can vary along a shoreline depending on the amount of water or height of breaking waves. Rip currents form as water tends to flow from regions of high setup (larger waves) to regions of lower setup (smaller waves), where currents converge to form a seaward flowing rip. 

\deleted[id=A1]{The speed of seaward rips can be very strong, reaching five miles per second, faster than an Olympic swimmer.}
\added[id=A1]{Detecting rip currents with machine learning (ML) is challenging because there are different types of rips, each with a different appearance. The three major factors that lead to different types of rips are the shape of the shoreline, the bathymetry, and hydrodynamic factors (e.g., wave height and direction, tides). The combination of these lead to rip currents with different visual signatures.  Rips may also either be transient or persistent in space and time.
Detecting rip currents pose unique challenges compared to detecting other objects such as cars, or
people, etc. Rips are amorphous without a well defined shape or boundary, and are ephemeral without
well defined temporal bounds.}

\added[id=A1]{Using an object detector, such as those included in a recent survey \cite{object_detection_survey}, would require a substantial training dataset for each type of rip current. To date, there are only two publicly available training datasets for rip detection: individually labeled frames \cite{de2021automated} and time averaged images \cite{maryan2019machine}, both of the same type of rip current.  There are no existing training dataset for other types of rip.}

By nature, rips always eventually flow seaward regardless of their visual appearance.
Therefore, we propose a novel hybrid approach that incorporates flow analysis with ML for rip detection, rather than relying on image colors.  We also propose encoding rips as locations with anomalous flow behavior, transforming the detection task into differentiating normal from anomalous behavior.  This greatly simplifies the collection of training data since only a single dataset is needed.

\deleted[id=A1]{Multiple factors determine the location and strength of rips, such as bathymetry, wave height
and direction, tide, and beach shape. Rip currents may either be transient or persistent in space and time. Rips frequently found at the same location over time usually indicate a relatively stable bathymetric feature such as a stable sandbar or reef or a solid structure such as a rocky outcrop, jetty, or pier.
%The bathymetric features result in wave breaking, and setup variations channelize rip current flow. 
Transient or flash rips on the other hand are usually independent of bathymetry and may move up or down the beach and may appear or disappear.}

There are two existing approaches to detecting rip currents from stationary videos. %(i.e. video obtained from a stationary source). 
The first approach uses the appearance of rip currents to detect them. This includes human experts analyzing time-averaged (Timex) video or running automated object detectors \cite{de2021automated, maryan2019machine, RipNet20, RipDet21} to detect rip currents. The second approach relies on flow analysis, where the flow behavior of ocean waves is analyzed to detect rip currents. Direction-based clustering of flow vectors \cite{philip2016detecting, mori2022flow} and timelines \cite{mori2022flow} placed parallel to the beach are used in this approach. While able to detect weaker rip currents or those where the appearance is not obvious, timelines \deleted[id=A-2]{do} require user input to specify their initial placement.  
% XXX note: we could also automate the shoreline detection, but for this paper we pursue hybrid method.
% However, these flow-based approaches are not automated but require user input during run time. 
 
\begin{figure}[tb]
    \centering
    \includegraphics[width=0.99\linewidth]{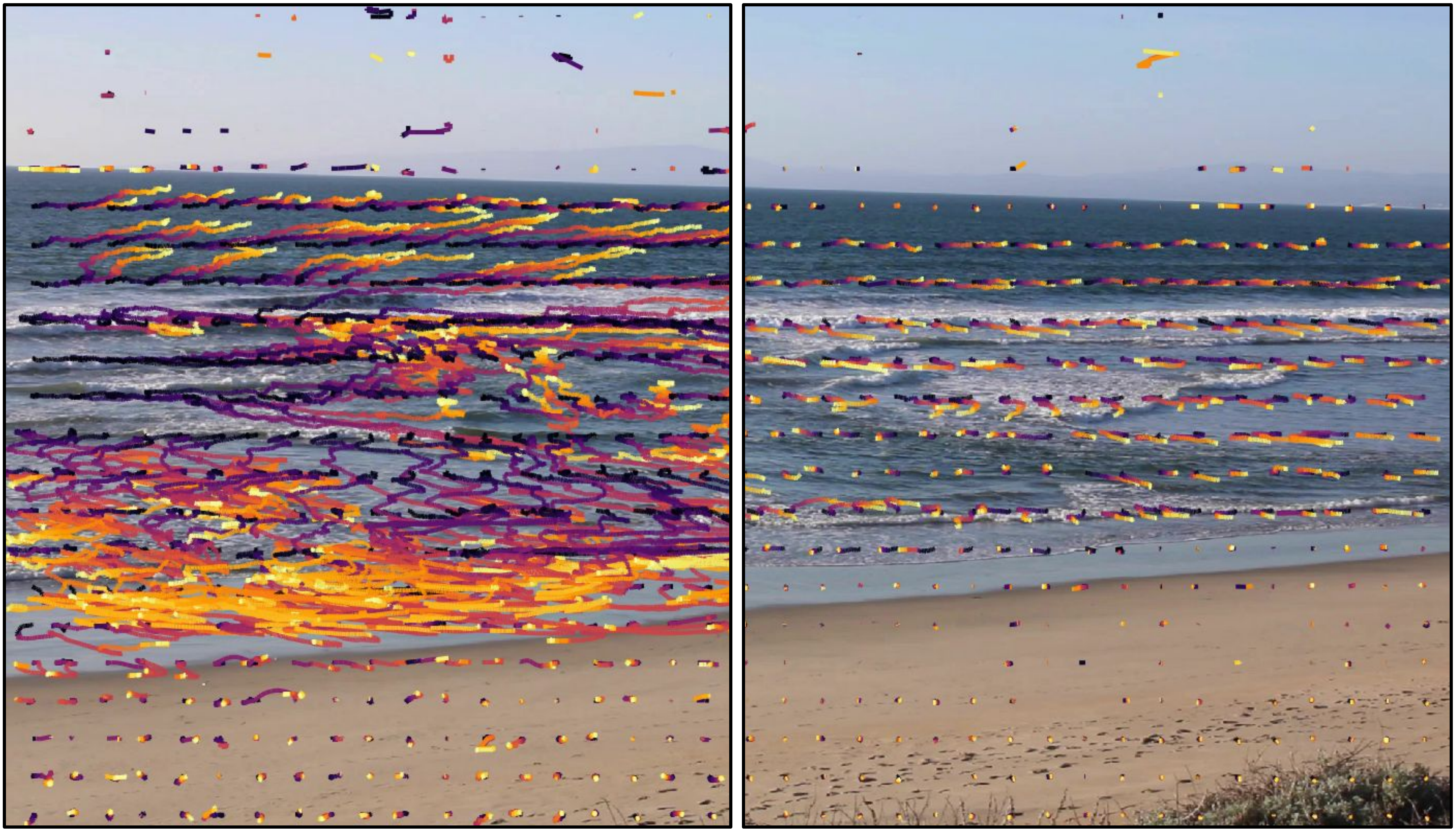}
    \caption{\textbf{Long pathlines get cluttered making it difficult for ML to learn:} Left pane shows longer pathlines integrated over the entire length of the video. Shorter pathlines, integrated over 900 time steps are shown in the right pane. Longer pathlines are much more cluttered and noisy, making it difficult for ML algorithms to learn its behavior. Notice the relatively large amount of noisy pathlines in the sky and the beach of the left pane.
    The pathlines are colored by age with yellow representing most recent. }
    %Most recent points on the pathlines are colored in a shade of  yellow.
    % \blue{the pathlines don't look long. can you change integration step for these 2 images by 10 or 100 fold. if dt was .0001, try .001 or .01 instead.}\red{sure let me see what i can do.}
    \label{fig:pathline_length_comparison}
\end{figure}

In this paper, we introduce RipViz, a fully automated, hybrid \added[id=A-2]{of} deep learning and flow analysis, feature detection method to find rip currents,  as shown in Figure \ref{fig:teaser}. We first obtain the time varying flow field from a stationary video using optical flow.
% this line seems weird
\deleted[id=A-2]{To analyze such a flow field, it is useful to know some basics about wave behavior.
Mid to long period waves usually come in groups of 2-5 waves on average,
followed by some period with lessened wave activity.  Short period waves are usually wind driven with short fetch and a less pronounced cycling of wave activity.
This cycle of wave activity is referred to as a wave set.
Our videos are generally 2-4 minutes long and sufficient to capture a wave set.} 
We use pathline sequences to capture the flow behavior. 
% this paragraph seems disconnected from the rest of the introduction
One could simply seed pathlines at every point and trace each of them for the duration of the video.  However, due to the quasi-periodic nature of wave activity, this simple approach produces too much clutter to be of much use as shown in Figure \ref{fig:pathline_length_comparison}. Furthermore, pathlines integrated over the full length of the video 
%are prone to be noisy as shown in the sky and the beach in \ref{fig:pathline_length_comparison}. 
accumulate more error, especially in noisy \added[id=A1]{real world} datasets.
Instead, we generate \textit{sequence of shorter pathlines} for each seed point.
%\blue{need a side-by-side figure here: one with full length pathlines, next to it, with shorter pathlines.}
By staggering the initiation of pathlines, our expectation is that \deleted[id=A-2]{pathlines} \added[id=A-2]{pathline sequences} outside the rip zone will behave differently.
For example, pathlines seeded in the surf zone where the waves are breaking will have large variations in their trajectories.  Pathlines seeded further out to sea, on the beach, and sky would be fairly still.  Pathlines seeded within the rip zone will have less variations in their trajectories yet will be different from the stationary ones.

% \begin{figure}[tb]
%     \centering
%   \subfloat[Small Pathline\label{fig:shorter_pathline}]{%
%       \includegraphics[draft,width=0.49\linewidth]{short_pathline.png}}
%     \hfill
%   \subfloat[Long Pathline\label{fig:longer_pathline}]{%
%         \includegraphics[draft,width=0.49\linewidth]{long_pathline.png}}
%   \caption{\red{work in progress} Comparison of a short and long pathlines.} 
%   \label{fig:shorter_and_longer_pathline_comparison} 
% \end{figure}

% To capture the flow behavior in the unsteady ocean flow, we seed pathlines across time at the same seed points to generate sequences of pathlines. 
In RipViz, we frame detecting rip currents as a flow anomaly detection problem. 
%\blue{RipViz employs a two stage processing of these pathlines to identify the rip zone.}
An LSTM autoencoder with a custom weighted loss function is used to learn the spatiotemporal features of pathline sequences for normal ocean flow (i.e. not rip currents). The trained LSTM autoencoder can predict anomalous pathline sequences (i.e. rip currents) during test time.
The origination points of anomalous pathlines are identified
and highlighted as a means of visualizing the rip zone. Our target users are general public who are not familiar with rip current dynamics nor visual
analytic systems. Hence, our design for the visualization output is to make it as simple and unambiguous
as possible.\\

%We feed sequences of pathlines generated from normal ocean flow (i,e., not rip currents) to train an LSTM autoencoder with a custom weighted loss function. 
% XXX not sure i understand the ramifications of the following sentences. maybe discuss these in more detail in later section.  We observed the reconstruction error of the trained LSTM autoencoder during run/inference time. We noticed that reconstruction error was high for anomalous pathline sequences (i.e., rip currents). The difference in the reconstruction error helped us select and visualize pathlines belonging to rip currents. \\

\noindent The main contribution of this paper is: 
\begin{itemize}
    \item A hybrid feature detection method that combines machine learning and flow analysis techniques to automatically find and visualize dangerous rip currents. 
\end{itemize}
  
\added[id=A1]{In order to realize this, the following innovations are necessary:}
\begin{itemize}
        \item \added[id=A-2]{For flow fields with a quasi-periodic behavior, such as ocean flow, working with a {\em sequence of shorter pathlines} is better than a single long  pathline. } 
        
        \item \added[id=A-2]{A weighted binary cross entropy, unlike the unweighted binary cross entropy used in previous works, is more effective when learning from sparsely distributed pathlines generated from ocean flow.  }
\end{itemize}
\deleted[id=A1]{We also introduced a {\em modified} FlowNet to detect rip currents (see Appendix A) and compared it with RipViz.}

% \begin{itemize}
%   \item  \deleted[id=A-2]{For certain flow fields, e.g. those with quasi-periodic behavior, working with pathline sequences is better than single pathlines. }
%     %(or splitting a single pathline into multiple pieces).
%      \deleted[id=A-2]{The pathline sequences allow the LSTM based autoencoder to learn about flow behavior more effectively.}
%     \item  \deleted[id=A-2]{We empirically determined the weights for the loss function for our driving application
%     and showed how such a weighted binary cross-entropy function can better learn from sparsely distributed pathlines.}
% \end{itemize}

% Papers related to rip currents : \cite{rip_types_castelle} \cite{brewster2019estimations} \cite{rip_stats_brighton} \cite{sea_bathing_stats_klein} \cite{rip_current_drowning_lushine}\\
% There is math equations / numerical methods can be used detect the boundary for vortices \cite{berenjkoub2020vortex}. However there is no such numerical methods to detect the boundary of rip currents. SVM+user input li et al. \cite{li2015extracting}

\section{Related Work \label{s:related_work}}

\vspace{2pt}\noindent\textbf{Streamline Selection: } Streamline selection and seed placement are well studied in flow visualization. Sane et al. \cite{sane20} provide a survey of the body of work over the last two decades.  These works share overlapping goals with research on streamline clustering \cite{carmo04}, and flow simplification \cite{lodha00}. Each aims to produce an uncluttered presentation of the flow field while still capturing the essential flow features and behaviors.  For example,
Marchesin et al. \cite{marchesin2010view} proposed dynamically selecting a set of streamlines that leads to intelligible and uncluttered streamline selection. Ma et al. \cite{ma2013coherent} used an importance-driven approach to view-dependent streamline selection that guarantees coherent streamline update when the view changes gradually. Yu et al. \cite{yu2011hierarchical} proposed hierarchical streamline bundles by producing streamlines near critical points without enforcing dense seeding throughout the volume. They grouped the streamlines to form a hierarchy from which they extracted streamline bundles at different levels of detail. Tao et al.\cite{tao2012unified} proposed two interrelated channels between candidate streamlines and sample viewpoints. They selected streamlines by taking into account their contribution to all sample viewpoints. However, the methods mentioned above use handcrafted features to define the feature representation of streamlines. Handcrafting features to represent pathlines in complex unsteady flow fields, such as ocean flow fields, is a challenging task. In contrast, the method presented in this paper learns complex feature representations without the need for handcrafted features.  

\vspace{2pt}\noindent\textbf{Deep Learning for Flow Visualization: } In recent years, the visualization community has worked with ML in two ways:
visualization to understand the ML model, and use of ML in visualization tasks.  On the latter, particularly for flow visualization tasks, \added[id=A-2]{Berenjkoub et al.\cite{berenjkoub2020vortex} used U-net, a deep learning neural network, to identify vortex boundaries.} 
\added[id=A-2]{Kim and G{\"u}nther \cite{kim2019robust} used a neural network to extract a steady reference frame from an unsteady vector field. } 
Han et al. \cite{han2018flownet} used an autoencoder-based deep learning model, FlowNet, to learn feature representations of streamlines in 3D steady flow fields which are then used to cluster the streamlines. \added[id=A-2]{They used longer streamlines that were integrated through the entire extent of the data. Furthermore, they used one streamline per seed point.}
\deleted[id=A-2]{Other researchers in the flow visualization community have also used deep learning methods to enhance visualizations.}
\deleted[id=A-2]{Berenjkoub et al used U-net, a deep learning neural network, to identify vortex boundaries.}
\deleted[id=A-2]{Kim and G{\"u}nther used a neural network to extract a steady reference frame from an unsteady vector field.}
The work presented in this paper uses \added[id=A-2]{a sequence of pathlines per seed point to learn the flow behavior from noisy unsteady 2D flow fields. This work uses} an LSTM autoencoder to detect anomalous pathline sequences from noisy unsteady 2D flow fields.

\vspace{2pt}\noindent\textbf{Rip Current Detection: } Traditional rip current detection generally involves in-situ instrumentation such as GPS-equipped drifters and current meters \cite{leatherman17,macmahan11}.  Remote sensing of rip currents is also possible with aerial imaging of the spread of fluorescein dye in the rip zone captured by drones, and the use of marine radar \cite{haller14,leatherman17}.
These require expensive equipment or a team of observers which makes them impractical for rip current monitoring or detection purposes.
However, with simple optical video capture such as from surf webcams, it is possible to detect certain rip currents using
a time-space (timestack) display or a simulated long time exposure images (timex) display
\cite{holman11}.
Timex images reveal rip locations as darker regions in the image that correspond to deeper channels in the bathymetry where water may flow seaward.  In contrast, incoming water associated with the breaking waves in the surf zone appear much brighter in timex images.
This type of rip is referred to as bathymetry controlled rips and are characterized by a quiet region in the rip channel that is flanked by breaking waves on either side.
Interpreting timex images require some domain expertise.
Maryan et al. \cite{maryan2019machine} used a Viola-Jones framework to train their model on timex images,
and indicate the location of the rips via bounding boxes.
\added[id=A1]{Likewise, Rashid et al. \cite{RipNet20, RipDet21} also used timex images but utilized a modified version of the Tiny-Yolo V3 architecture.
Similarly, Ellis and McGill \cite{OptFlowTimex} used timex images in conjunction with environmental information such as tides, wave height, and period to cluster offshore movements to rip currents.
However, their method produced false positives in situations where non-water objects such as surfers, paddle boarders, etc., are also moving offshore.}
Rather than working with timex images, 
de Silva et al. \cite{de2021automated} trained a Faster R-CNN model \added[id=A-2]{with a accumulation buffer} to detect bathymetry rips using individual frames of the video.
\deleted[id=A1]{Both}\added[id=A1]{All} of these methods detect \added[id=A1]{bathymetry }rips based on the appearance of the sea state. However, in instances where appearance is different or weak, these methods fail to detect rip currents.
%There are two main methods for visualizing rip currents using appearance or flow behavior. Among the appearance-based methods, observing time average images or employing object detectors \cite{de2021automated} were widely used to detect rip currents. However, applying flow visualization methods to visualize rip currents has recently gained traction by reconstructing the vector field using optical flow. 

\begin{figure*}
    \centering
    \includegraphics[scale=0.35]{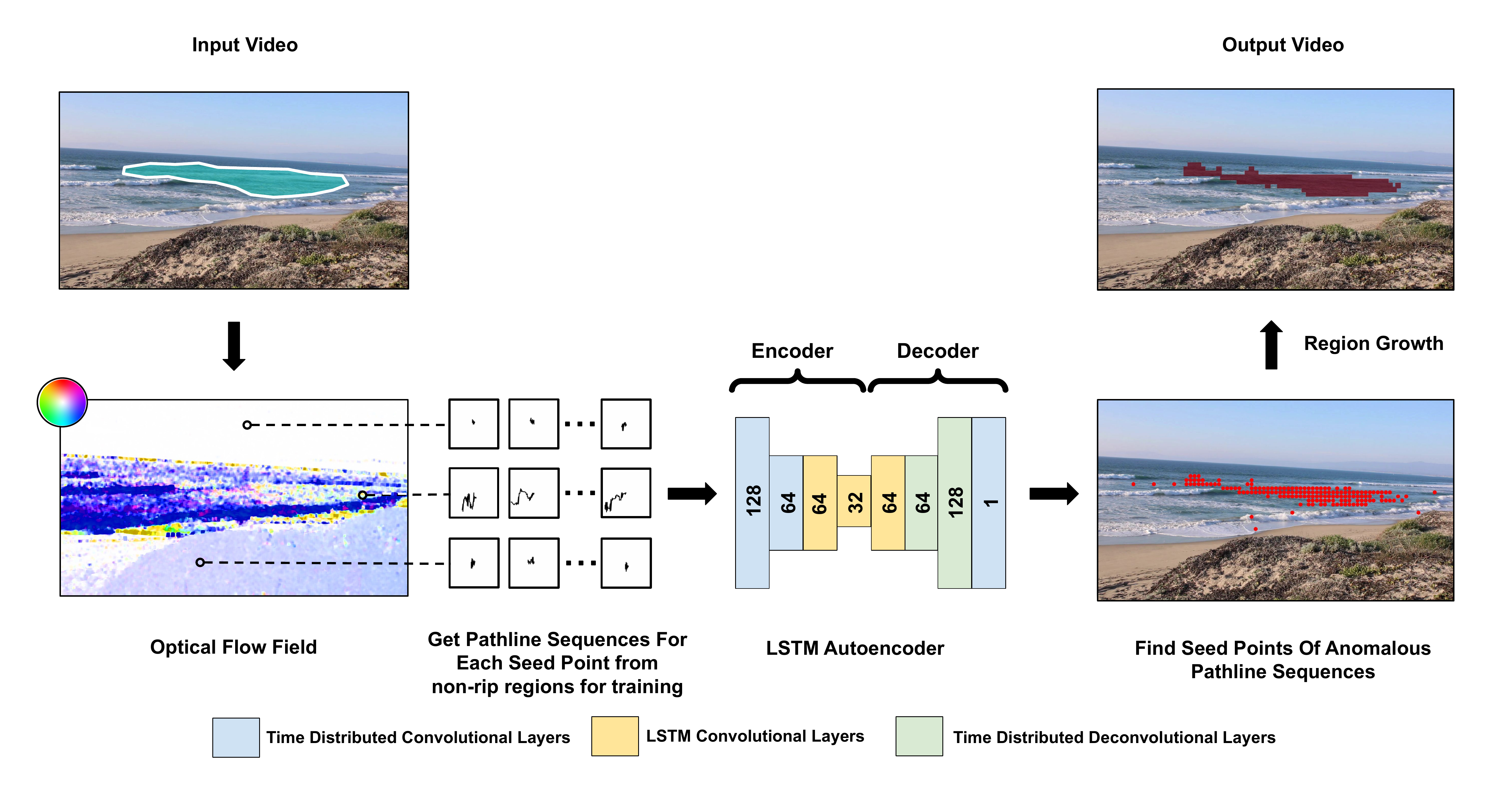}
    \caption{\textbf{RipViz pipeline:} First, an optical flow field was generated from the input video (magnitude and direction of velocities are mapped to value and hue, respectively.) Then pathline sequences were generated for each seed point in non-rip regions (only three seed points are shown) where a new pathline is seeded at every frame. Finally, an LSTM autoencoder was trained with these pathline sequences generated from non-rip regions. For rip detection, the trained LSTM autoencoder is used to detect anomalous pathline sequences by regularly seeding the video frame. We then applied a region growing algorithm to the anomalous points to find the anomalous region, while filtering out singleton seeds.}

    \label{fig:full_pipeline}
\end{figure*}

% \begin{figure*}[tb]
%     \centering
%     \includegraphics[scale=0.35]{figures/pipeline/pipeline_test.pdf}
%     \caption{\textbf{Testing Pipeline of RipViz:} First, an optical flow field was generated from the input test video (magnitude and direction of velocities are mapped to value and hue respectively).  Then pathline sequences were generated for each seed point (only three seed points are shown) where a new pathline is seeded at every frame. Pathline sequences were fed to the trained LSTM autoencoder. Once, anomalous pathlines were detected their seed points were rendered onto the video (note that only a few anomalous seed points are shown for clarity).  Normal pathlines and their seed points were not rendered onto the video. }
%     \label{fig:testing_pipeline}
% \end{figure*}

% \begin{figure*}[tb]
%     \centering
%     \includegraphics[scale=0.4]{figures/LSTM-autoencoder/RIPVIZ3.pdf}
%     \caption{Network architecture of RipViz. We use LSTM convolutional layers to learn the spatiotemporal features of the pathline sequences. \added[id=A-2]{THIS FIGURE WILL NE REPLACED BY FIGURE 3 and 4.}}
%     \label{fig:LSTM-autoencoder}
% \end{figure*}

An alternative approach is to detect rips based on the observed behavior. Philip and Pang \cite{philip2016detecting} obtained an unsteady flow field from the video and hypothesized that the rip current is directly opposite the dominant flow in the vector field due to the incoming wave motion. They grouped vectors based on the direction and magnitude to visualize the rip current. Mori et al.\cite{mori2022flow} used a similar approach to group vectors and improved the visualization of the rip current by mapping direction to color and magnitude to hue. In the same work, Mori et al.\cite{mori2022flow} also used timelines to visualize rip currents. They placed timelines parallel to the beach and observed its shape as it gets dragged by the rip current. 
The work presented in this paper also uses optical video of the sea state to detect rip currents.
The underlying approach is also based on the flow behavior, and hence not constrained to detecting bathymetry rips.  
\added[id=A1]{However, the presented methodology in this paper is the first to propose combining ML and flow analysis to detect different types of rip currents.  Because detection is based on treating flow behavior in rip currents as anomalous, the task of collecting and labeling training data for an ML model is unified and simplified.  In contrast, a standard ML rip detection model would require a training data set for each type of rip current -- a costly and time consuming process.}

\deleted[id=A1]{The methodology is a combination of flow analysis and machine learning.}

\vspace{2pt}\noindent\textbf{Autoencoders: } Autoencoders are a type of neural network that can learn object representations without any supervision \cite{autoencoder_survey_01, autoencoder_survey_02}. Autoencoders are trained with a loss function that compares the input and reconstructed output by using a reconstruction error. Long short term memory (LSTM) autoencoders are a specialized type of autoencoder that can learn from sequential data. These types of autoencoders are equipped with LSTM layers that can learn how data is temporally related.  Furthermore, autoencoders are also used as anomaly detectors\cite{anomaly_ribeiro2018study}. For anomaly detection, autoencoders are first trained with {\em normal} data. 
When presented with {\em anomalous} data,
the trained autoencoder will produce a high reconstruction error. We exploit this property to detect anomalous flow behavior.

The flow visualization community has used autoencoders to find feature representations of objects that then can be used for clustering. In their work, Han et al. \cite{han2018flownet} proposed to learn feature representation of streamlines and stream surfaces of 3D steady flow fields by using an autoencoder with a binary cross-entropy function. They then further reduced the dimensionality of the features and used those to generate clusters. \added[id=A-2]{However, their method does not learn how pathlines are temporally related. Additionally, their unweighted binary cross entropy loss function does not account for sparsely distributed pathlines.} \deleted[id=A-2]{In our work, we frame rip current detection as an anomaly detection problem.} \added[id=A-2]{In our work, w}e use an LSTM autoencoder with a custom weighted binary cross-entropy loss function, to learn spatiotemporal behavior of \added[id=A-2]{sparsely distributed} pathline sequences from ocean \added[id=A-2]{scenes}. 

%We frame the problem of rip current visualization as an anomaly detection problem. We train the LSTM autoencoder with sequences of pathlines originating from the same seed point from normal ocean flow (i.e., not rip currents). The trained LSTM autoencoder will generate a high reconstruction error when presented with anomalous pathline sequences (i.e., rip currents). We use this difference in reconstruction error to determine pathline sequences that belong to the rip current. %\red{ADD MORE INFO ABOUT AUTOENCODERS, SPELLOUT LSTM}

\section{RipViz \label{s:method}}
Identifying rip currents in complex and chaotic ocean flow is challenging. We first reconstructed a 2D unsteady flow field using optical flow from ocean videos. Our method captured the ocean flow characteristics by using short pathline sequences which are then represented as stacks of binary images. We used an LSTM autoencoder to learn spatiotemporal features of these pathline sequences. We trained our model with pathline sequences of normal ocean flow using a custom weighted binary cross-entropy function, suited for learning pathline sequences of ocean flow. We then used this trained LSTM autoencoder to identify pathline sequences of abnormal ocean flow or rip currents. We visualized the rip currents by projecting the seed points of these abnormal pathline sequences back onto the video frame \added[id=A1]{and growing a transparent red region around these points}. 

\subsection{Flow Field Reconstruction}
An unsteady 2D flow field is obtained from the stationary video using optical flow.
%We reconstructed the flow field from stationary videos using optical flow.
Many optical flow algorithms use the relative motion of neighboring pixels between consecutive frames in the video to calculate the local flow. We use Lucas-Kanade \cite{lucas1981iterative} sparse optical flow function in the OpenCV library \cite{bradski2000opencv} to trace pathlines. We verified each integration step by comparing the forward and backward integration of the flow field.  
%\blue{why not farneback?  how many points are you use with LK i.e. talk about sampling resolution.
%immediate output of LK/farneback is a flow field.  say something about how pathlines are calculated from that ... by the same LK call in opencv?}

\subsection{\added[id=A1]{Sequence of Pathlines} \deleted[id=A1]{Pathline Representation} \label{s:pathline_representation}}
Each pathline is represented by a 1D vector, $\textbf{p} = \{x_1, y_1, \cdots, x_n,y_n\}$, where $(x_i, y_i)$ is a point in the coordinate system of the video frame and $n$ is the length of the pathline. 
%\blue{does n correspond to the number of frames?}
For the LSTM autoencoder to learn about the pathlines, we transformed these pathlines into their own 2D pathline-centric coordinate system. To do this, each pathline is represented by an
$L \times L$ binary image $\textbf{I}$, where each point in $\textbf{p}$ is translated \deleted[id=A-2]{by $-(x_1, y_1)$} to center the seed point in the binary image.
%\blue{if 2 points are more than 1 pixel from each other, do you interpolate and fill the inbetween pixels with 1?}
%initialized with zeros. We translated the seed point of the pathline, point $(x_1, y_1)$ of $\textbf{p}$,  to the center of the binary image. If the pixel in $\textbf{I}[x_i, y_i]$ is occupied by the pathline, we updated the pixel value to 1; otherwise, we left it as 0. 
For longer pathlines that extend beyond $L \times L$,
%the binary image was too small to contain them. In such cases, 
we increased the size of the binary image to accommodate 
these longer pathlines. We then resized all of these larger binary images down to $L \times L$. \deleted[id=A-2]{by using bilinear interpolation.}
The reason why all pathlines were not simply centered then resized to $L \times L$ is that we need to differentiate pathlines that barely moved from their initial seed position versus those that actually travelled beyond $L \times L$.
Also note that by centering, the 2D representation of the pathlines are now location agnostic.
This combination allows us to compare pathlines from different parts of the video to identify those with similar behavior.
As noted earlier, tracing pathlines over the entire length of a video of quasi-periodic motion derived from noisy optical flow calculations result in unusable cluttered flow representations.
Instead,
we reseeded pathlines at the same seed point over regular time intervals to generate pathline series of length $S$. We represented each pathline sequence as a stack of binary images of size $S \times L \times L$ associated with each seed point.  
%\blue{specify what S and L are and provide some justification for their value -- maybe because of the video resolution and periodicity of waves.  if you're planning to talk about this later, include with a forward reference e.g. Parameter selection for S and L are discussed in Section blah.}

% Optical flow papers : \cite{lucas1981iterative}

\subsection{LSTM Autoencoder}

We used an LSTM autoencoder to learn from the pathline sequences. The LSTM autoencoder consisted of two components, an encoder, and a decoder, as shown in Figure \ref{fig:full_pipeline}. The encoder learns the spatiotemporal features of the pathline sequence. The decoder reconstructs the pathline sequence by using the learned spatiotemporal feature representation. 

We specify the input layer of the LSTM autoencoder to expect pathline sequences of shape $S \times L \times L$. The first and second layers of the encoder are time-distributed 2D convolutional layers. These layers consist of $128$ and $64$ convolutional filters, respectively. These layers process each pathline of the sequence separately and learn the spatial features of each pathline. The third and fourth layers are 2D convolutional LSTM layers. Each layer consists of $64$ and $32$ convolutional filters, respectively. These convolutional LSTM layers process the sequence of pathlines together and learn the temporal features of the pathline sequences. The first layer of the decoder is a convolutional LSTM layer with $64$ convolutional filters. The second and third  layers of the decoder are time-distributed deconvolutional layers with $64$, and $128$ respectively. The last layer of the LSTM autoencoder is a time-distributed convolutional layer with $1$ convolutional filter. Each layer, except for the input and the output layers, is followed by a normalization layer \cite{ba2016layer}. The input volume has no padding; therefore, we set the stride of all layers to $1$. The autoencoder outputs a sequence of pathlines of shape $S \times L \times L$. For the hidden layers, we use the ReLU activation \cite{RELU}. For the output layer, we use the sigmoid activation. 

In comparison to recent autoencoder based stream line selection methods, our method not only learns the spatial features of each pathline but also learns how each pathline is temporally related to other pathlines in the same pathline sequence. Learning these spatio-temporal features allows our method to better learn about the quasi-periodic nature of the chaotic ocean flow. 
% \blue{perhaps, space permitting, say something about how this is similar/different from flownet.}

\subsection{\added[id=A-2]{Weighted} Loss Function \label{s:loss_function}}
We trained the LSTM autoencoder by minimizing the difference between each training sample against the corresponding model prediction. The loss function used to compute this difference is a weighted binary cross-entropy loss function. Since each training sample is a pathline sequence represented as a stack of binary images of size $S \times L \times L$, we treat each sample as a binary volume where $p_i$ is $1$ when the pathline crosses that voxel and $0$ otherwise. $\hat{p_i}$ is the corresponding value from the predicted volume. The loss is calculated over all voxels (i.e., $N = S \times L \times L$ ).    
\begin{equation}\label{eq:weighted_loss_function}
{\mathcal{L} =  - \frac{1}{N} \sum_{i=1}^{N} [w_1 \cdot p_i \log \hat{p_i} + w_0 \cdot (1-p_i)\log (1 - \hat{p_i})] } 
\end{equation}

\noindent $w_1$ is weight assigned to voxels when $p_i=1$. $w_0$ is weight assigned to voxels when $p_i=0$. 
In contrast, previous autoencoder based neural network architectures used in flow visualization tasks used an unweighted version of the same loss function (i.e., $w_1 = w_0 = 1$ ). We found that a weighted binary cross-entropy function is better suited for our application domain for reasons discussed in Section \ref{s:hyper_parameters}.
% \blue{explain why this is significant.  also since $p_n$ is binary, only 1 of the terms is non-zero. what are the values of $w_0$ and $w_1$?  also, subscript i is not used in the equation -- should $p_n$ be $p_i$?}\red{updated $i$ to $n$, $w_0$ and $w_1$ will discuss in the hyper-parameter tuning section}

\subsection{Detecting and Visualizing Rip Currents \label{s:viz_rips}}

During inference/test time, we calculated the reconstruction error for each pathline sequence by using the loss function defined earlier.
%in Section  \ref{s:loss_function}. 
If the error is larger than threshold $T$, we labeled those pathline sequences as anomalous (i.e., rip currents). Otherwise, we labeled the sequences as belonging to normal \deleted[id=A1]{ocean} flow. We discuss how $T$ was selected in Section \ref{s:hyper_parameters}.
%\blue{talk about how T is selected, or do a forward reference if you're planning to talk about this later}
\added[id=A-2]{Once the anomalous pathline sequences were found, the corresponding seed points were connected by using a region growth algorithm to generate a region.}
\added[id=A1]{Isolated seeds without neighbors are discarded.}
 To visualize the rip zone,
We projected \added[id=A-2]{this region} \deleted[id=A-2]{the seed points of anomalous pathline sequences} back onto the frames in the video as shown in Figure \ref{fig:teaser}.
%\blue{refer to some figure, talk about color coding.}

\subsection{Network Training \label{s:network_training}}
 We implemented our neural network in Tensorflow by using the Keras application programming interface (API)\cite{chollet2015keras}. We trained the network using an NVIDIA Tesla A100 graphical processing unit (GPU). 
 %We used a similar approach as Han et al. \cite{han2018flownet} when training our deep neural network. 
 In the training process, we initialized parameters in all layers of the neural network using a normal distribution $\mathcal{N}(\mu, \sigma^{2})$, where mean $\mu = 0$ and variance $\sigma^{2}=0.01$. 
 %We set the $random\ seed =$ $1234$ in the Keras API for reproducibility.
 We applied the Adam optimizer [38] to update the parameters with a learning rate of $10^{-6}$. %\blue{is there some units to this "rate"?}.
 We used the minibatch size of \deleted[id=A-2]{$4$} \added[id=A-2]{$10$} and trained our model with $100$ epochs. %\blue{what's the termination criteria, small delta from previous epoch, or reached some predefined performance, or some predefined amount of time, or ..}

\section{Results and Discussion \label{s:results}}

\added[id=A1]{We describe our dataset, provide an analysis of the components and parameters used in RipViz, and present comparisons with other methods.}

\subsection{Dataset}

Our dataset consists of $55$ stationary videos. Each video is $1.5$ to $4$ minutes long and of size $1920 \times 1080$ pixels. We collected the data from Salinas, Marina, and Davenport beaches in California in winter of 2022. We used either a tripod-mounted Canon EOS Rebel T7 DSLR camera or a Samsung Android phone to acquire the data. We chose $4$ videos to generate $16000$ non-rip pathline sequences for training our model by using the seeding strategy discussed in Section \ref{s:hyper_parameters}. \added[id=A-2]{Our validation dataset consists of 12000 pathlines generated from three videos. The remaining videos were used for testing. Note that our training data does not include examples of each of the many types of rip currents, only examples of normal ocean flow. This greatly simplifies collection of training data.} \added[id=A-2]{Each video was labeled under the guidance of the rip current expert. Pathlines seeded in the rip current area were labeled as ``rip" and others as ``non-rip".}

\added[id=A1]{The framing of these videos, which includes the distance of camera to the water and the focal length or zoom factor of the lens, is meant to be representative of surf webcams (e.g. surfline.com) that might be useful for rip current detection.  Parameters such as integration length $n$ and binary image size $L$ are based on such framing.  Sensitivity of these parameters on different framings are discussed in Section \ref{s:hyper_parameters}.
}

\added[id=A1]{
Note that we assume a stable video source and hence did not perform any video stabilization.
Wind can cause slight movements of the camera, but also movements of grass, clouds, etc.
Also,
video processing seldom work with raw video but rather on compressed video. Deriving the optical
flow field on compressed video may also introduce some motion artifacts especially in highly compressed regions
such as the sky or empty sandy beaches.
RipViz handles these type of motion artifacts better than existing methods for our dataset.}\\

\begin{figure}[tb]
    \centering
    \includegraphics[width=0.99\linewidth]{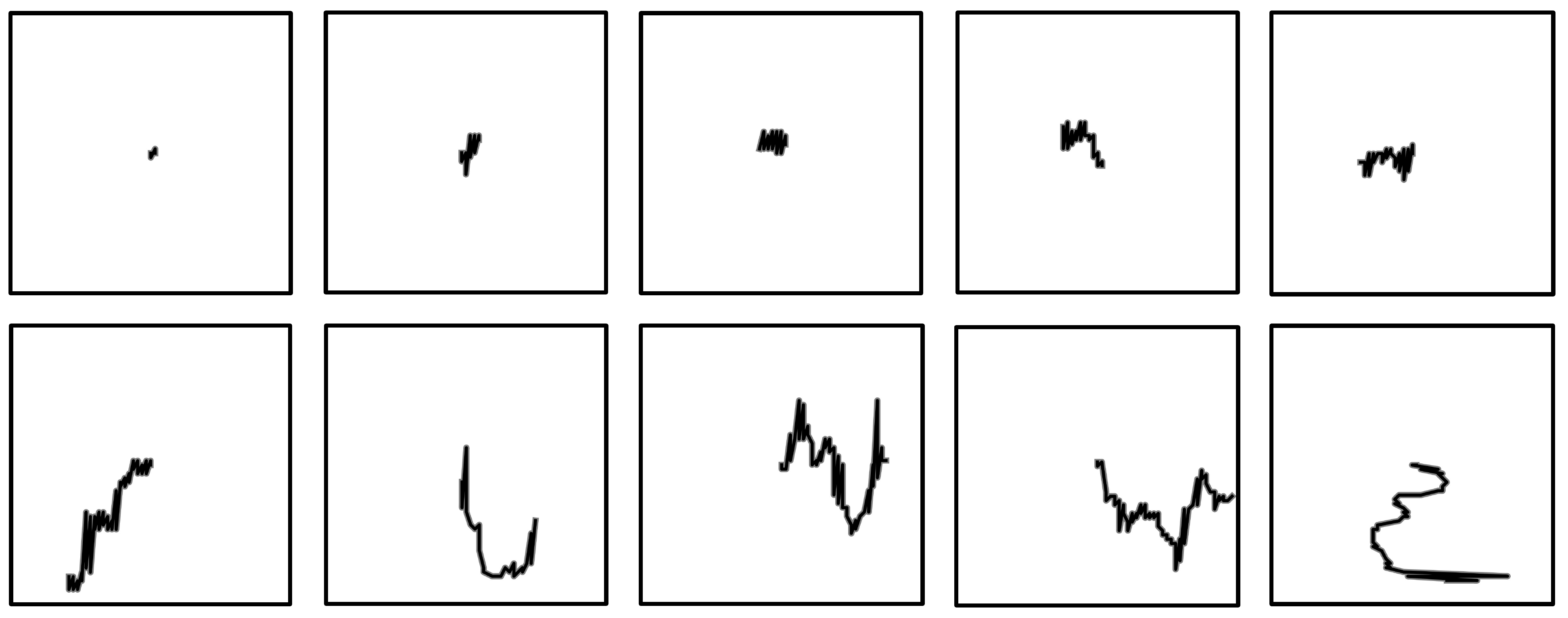}
    \caption{\textbf{Sparsely distributed pathlines in binary image:} This figure shows a collection of pathlines from our data.  Notice the imbalance in $0$s (white) and $1$s (black) in the binary images. \added[id=A-2]{The weighted loss function, as described in Equation \ref{eq:weighted_loss_function} allows the autoencoder to learn with imbalanced data.}} %\red{will update with a high res figure soon}}
    \label{fig:short_and_long_pathline_comparison} 
\end{figure}

\subsection{Analysis of RipViz components \label{s:hyper_parameters}}

The RipViz method contains two important changes from prior methods: the use of a weighted loss function, and the use of a sequences of pathlines rather than a single pathline. Without these changes we found the ML fails to learn ocean flow. In addition, the method has parameters like detection threshold, $T$, and binary image size $L$ which are likely dependent on our specific application domain. In this section we analyze each of these factors, showing that our changes are necessary, and providing the method by which we determined parameters.

% \vspace{2pt}\noindent\textbf{Amount of training data: }\red{work in progress}\\
\vspace{2pt}\noindent\textbf{Threshold $T$:} Threshold $T$ is used to filter anomalous pathlines (i.e, rip currents) based on the reconstruction error as discussed in Section \ref{s:viz_rips}. In order to find the optimal threshold $T$ we calculated $F_1$ score at varying threshold values in a subset of our data. $F_1$ score is defined as,

\begin{equation} \label{eq:f_1_score}
    F_1 = \frac{2}{\frac{1}{recall} + \frac{1}{precision}} = \frac{2}{\frac{FNs+TPs}{TPs} + \frac{FPs+TPs}{TPs}}
\end{equation}

\noindent If a detected pathline falls within the expert annotated boundary of the rip current, then it's counted as a TPs (true positive). Otherwise, it's considered an FP (false positive). Suppose a pathline originating within the rip current is not detected, then it's an FN (false negative). The range of the $F_1$ score is between $0$ and $1$. If most pathlines fall within the rip current boundary, the score will be closer to $1$, otherwise closer to $0$.
We found that threshold, $T = 1.0 $ 
%\blue{sure it's not 0.13??} % In the new model the threshold I am using is $1.0$. This was different from the one I reported for the vis paper 
%\blue{$T = 0.125$} 
produced the highest $F_1$ score 
% as shown in Figure \ref{fig:param_tune_threshold}
, and was used as $T$ for the experiments in this paper.    
% In our context, given a voxel in the binary volume, it is a true positive/false positive/false negative if the value of ground truth is 1/0/0 and the possibility predicted by FlowNet is greater/greater/less than 0.5. Ranging between 0 and 1, F1 score defines the similarity between the original and the predicted objects. If F1 score is closer to 1/0, it indicates that the predicted object is more/less similar to the original object.

% \begin{figure}[tb]
%     \centering
%     \includegraphics[width=0.9\linewidth]{figures/threshold/figure_donnie (1).pdf}
%     \caption{$F_1$ score at varying threshold $T$ values. Notice that    $T=0.013$ was optimal. \added[id=A-2]{WILL MOVE TO SUPP MATERIALS: SEEMS NOT RELEVANT TO THE PAPER}}
%     \label{fig:param_tune_threshold}
% \end{figure}

\vspace{2pt}\noindent\textbf{Weights of the loss function $w_0$ and $w_1$: }
We reconstructed the flow field from videos using optical flow. 
%Since some of the seed points fell on the still parts of the video frame, such as the beach and the sky, we noticed that pathlines traced from those seed points were relatively short (though not perfectly still due to minor shaking from wind and noise from estimations of the optical flow map). 
As discussed in Section \ref{s:pathline_representation}, all pathlines are represented in a binary image $\textbf{I}$.
%We sawthat smaller pathlines only spread within a fewer number of pixels compared to larger pathlines, as shown in Figure \ref{fig:short_and_long_pathline_comparison}. 
We observed that some pathlines are short and do not extend far from the seed point, while other pathlines are longer and travel farther from the seed point, as shown in Figure \ref{fig:short_and_long_pathline_comparison}.
We needed the model to learn the distribution of these shorter and longer pathlines because it allowed the model to learn the \textit{normal} flow behavior in an ocean scene. However, this leads to an imbalance in the number of voxels with 1s and 0s, especially for shorter pathlines. For the deep learning model to learn effectively about these small features/pathlines, we needed to increase the weight of the loss function when the target probability ($p_i$) is $1$. 
% \blue{what are you referring to by "target probability is 1"? do you mean where the pixel is 1, or shorter pathlines?}
%\blue{is there another term to use instead of ground truth pixel value?  how about just pixel value?}
The loss function, as described in Section \ref{s:loss_function}, penalizes the model more for making mistakes when predicting pixels with the target probability $1$ in the training process. Using the weighted binary cross-entropy function made the model better learn the distribution of shorter from longer pathlines. In contrast, in FlowNet \cite{han2018flownet}, the streamlines were well spread out across their input volumes, making the use of a weighted loss function unnecessary.
% \blue{so model just distinguish between short vs long -- and not really rip vs non-rip or normal vs abnormal?}

% \blue{possibly rephrase as: .. made the model better differentiate shorter from longer pathlines(?)}
% \blue{question: is the loss function really just trying to distinguish between short (beach/sky) vs long (ocean) pathlines?  doesn't it also need to distinguish pathlines in normal ocean parts vs rip zone?  the normal ocean pathlines are longer too..}

\begin{figure}[t]
    \centering
    \includegraphics[width=0.95\linewidth]{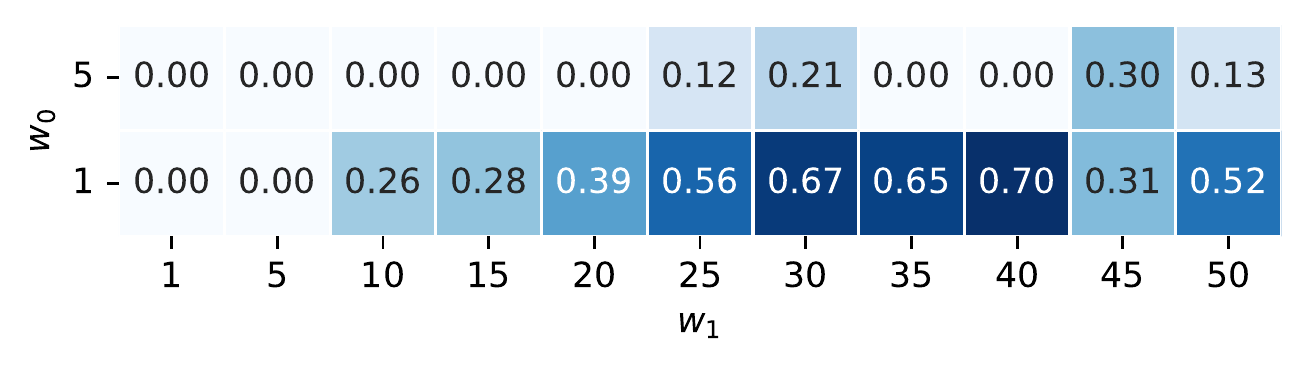}
    \caption{\textbf{$F_1$ Scores for different combinations of $w_0$ and $w_1$:} Notice that the method does not converge when $w_0$ and $w_1$ are balanced. \deleted[id=A1]{It is necessary to use increase $w_1$ for the model to converge.}\added[id=A1]{The highest $F_1$ score of 0.70 is achieved when $w_0$ is set to 1 and $w_1$ is set to 40.} }
    \label{fig:heatmap}
\end{figure}

% \red{work in progress} In our experiments we found that using $w_0 = w_1 = 1$ lead to an unoptimized model for the pathlines in our application domain.  However $w_1 = 10$ or $w_1 = 50$ while keeping $w_0 = 1$ lead to more optimized model.
% % \blue{reviewers are going to question why 10 and 50.  is it fair to say w1 = [10..50] works well?
% % they can guess what happens when w1 is less than 10, though making that explicit helps.
% % but happens when w1 > 50?  explaining that will also help.  if you have evidence e.g. figure from experiments,
% % that would be best.}
To find the optimal $w_0$ and $w_1$ for our application domain,  we randomly choose a small subset of pathline sequences, some short and some long, to train models at different combinations of $w_0$ and $w_1$. We trained each model for $100$ epochs, using the same training conditions specified in Section \ref{s:network_training}. For each trained model, we observed how many voxels were correctly recreated. If the predicted probability for a voxel with target probability of $1$  
%\blue{whatis a pathline voxel? also, why voxel?} 
is greater than $0.7$, we marked that voxel as TP; otherwise, it is considered to be FN.
%\blue{rephrase next sentence} 
If the model predicted pathline voxels in areas with no pathline, we marked that as FP. We calculated the $F_1$ score for each model with different weight combinations using equation \ref{eq:f_1_score}. We observed that non-uniform weighting is necessary, with the highest $F_1$ score generated when $w_0 = 1$ and $w_1 = 40$ as shown in Figure \ref{fig:heatmap}. Note that using unweighted cross-entropy, as in previous work, is the equivalent of setting $w_0=w_1=1$. In this condition we found that the model completely fails to learn ocean flow. 

In some instances where the model did not converge, TPs were $0$, which resulted in an undefined $F_1$ score. In such instances, we followed the default reporting convention of Scikit-learn python library \cite{pedregosa2011scikit} and reported those $F_1s$ as $0.00$.

%% when TPs = 0, F_1 score is undefined. But according to some (see link below) its okay to put 0 in those instances. 
%%https://github.com/dice-group/gerbil/wiki/Precision,-Recall-and-F1-measure

\begin{figure}[t]
    \centering
    \includegraphics[scale=0.9]{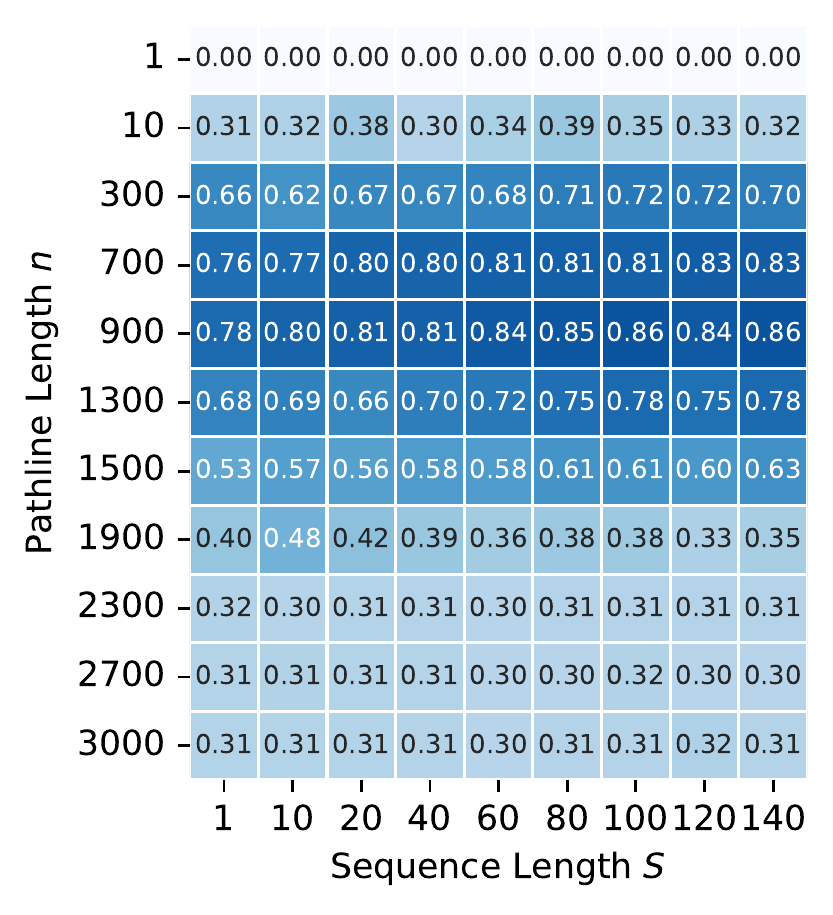}
    \caption{\textbf{Pathline length ($n$) vs sequence length ($S$)}: $F_1$ scores for different combinations of $n$ nd $S$.
    The first column represents traditional pathlines, while subsequent columns are for sequence of pathlines. Notice that increasing the pathline length resulted in higher accuracy up to an optimal length; beyond this, the accuracy decreased. Also notice that using a sequence of pathlines over a single pathline per seed point produced even more accurate results. We found that $n=900$ and $S=100$ were optimal for our data set.}
    \label{fig:heatmap_n_S}
\end{figure}

\begin{figure}[t]
    \centering
    \includegraphics[width=0.95\linewidth]{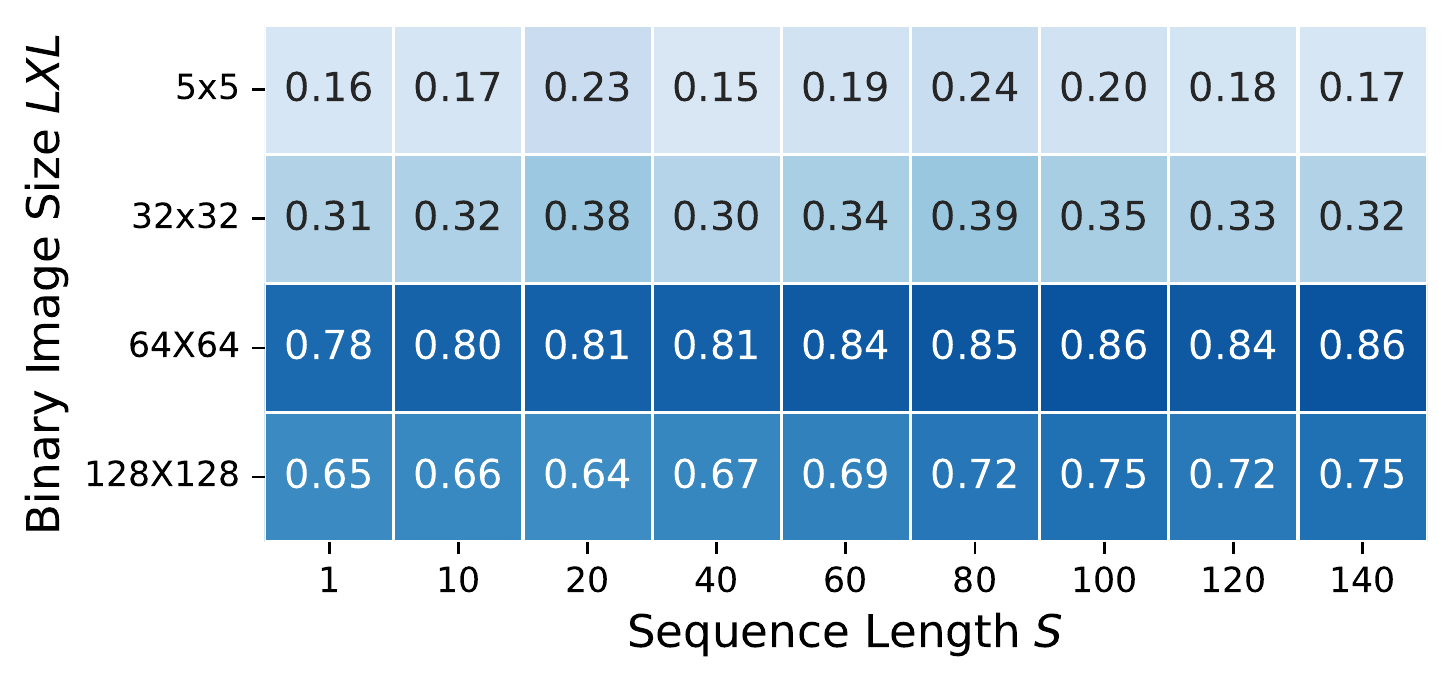}
    \caption{\textbf{Binary image size ($L \times L$):} Notice that a binary image size of $64 \times 64$ produced the highest $F_1$ score with least amount of computation. }
    \label{fig:binary_image}
\end{figure}

\vspace{2pt}\noindent\added[id=A-2]{\textbf{Pathline Length $n$ and Sequence Length $S$: } In RipViz, we use pathline sequences to learn flow behavior. In order to find the optimal pathline length, $n$, and sequence length, $S$, for our application domain, we trained multiple models while changing $n$ and $S$, and observed the $F_1$ score as shown in Figure} 
\ref{fig:heatmap_n_S}. \added[id=A-2]{We noticed two patterns from this experiment. First, increasing the pathline length resulted in a higher accuracy up to an optimal length; beyond this, the accuracy decreased. Second, using a sequence of pathlines instead of a single pathline per seed point produces more accurate results. We found that $n=900$ and $S=100$ were optimal for our data set. }

\vspace{2pt}\noindent\deleted[id=A-2]{\textbf{Length of pathlines $n$: } In RipViz, we use pathlines to capture the flow behavior. As discussed in Appendix B, longer pathlines accumulate more error in noisy datasets such as ours. Also, longer pathlines tend to clutter compared to shorter ones, as discussed in the introduction. In our experiments, we tried to find the shortest possible pathline that could sufficiently capture the flow behavior in our application domain. In order to find the optimal $n$, we trained four models by varying $n$ with our training data. We trained our model with two short pathlines $n=10, 100$, and noticed that short pathlines are sub-optimal at capturing flow behavior. We also noticed that very long pathlines $n=5000$ were too cluttered to accurately capture the flow behavior. We calculated the $F_1$ score using equation \ref{eq:f_1_score}. We found that $n = 900$ produced sufficiently high $F_1$ as shown in Table, and was used as $n$ for our experiments. }

\vspace{2pt}\noindent\deleted[id=A-2]{\textbf{Length of the pathline sequence $S$: } 
Our model is trained on pathline sequences of length $S$ as discussed in Section \ref{s:pathline_representation}. We noticed that longer pathline sequences are better at capturing the flow behavior at a seed point. We experimented with pathline sequences of varying lengths as shown in Table . We found that pathline sequences of $100$ are sufficient in capturing the flow behavior in our data. }
%[Also note that we hypothesize that even longer pathlines sequences are even better at learning the flow behavior, however due to memory limitations in the $GPU$ used, we are currently unable to do experiments with $S$ more than $100$]. 
% \blue{maybe put back the commented sentence above if that's really the case.  also, is it a coincidence that all 3 tables have 0.85 as best score?}

\vspace{2pt}\noindent\textbf{Size of binary image $I = L \times L$: } We represented each pathline in a binary image as discussed in \ref{s:pathline_representation}. We trained \added[id=A-2]{several} \deleted[id=A-2]{four} models with our training dataset where binary images were of different sizes as shown in Figure \ref{fig:binary_image}. We compared each model by calculating the $F_1$ score, similar to how $F_1$ score was calculated for tuning Threshold $T$. Our experiments indicated that small binary images ($L \times L=5 \times 5$) cannot sufficiently capture the variations of pathline movements,  resulting in a lower $F_1$ score. We also found that larger binary images ($L \times L=128 \times 128$) also resulted in a slightly lower $F_1$ score. Furthermore, larger binary images resulted in a high memory usage and a longer training time.  We found that $L \times L=64 \times 64$ is sufficient at capturing the variations of pathlines resulting in a high $F_1$ score. In our experiments we used $L \times L=64 \times 64$ as the size of the binary image $I$. 
% \blue{why start with 5x5? maybe replace that with 32x32, or include a 16x16 as well?}

%\blue{is 128x1282 not as good?  may want to show computational cost to justify going with a smaller size..}
%[Memory limitations are also a problem for this.]

\vspace{2pt}\noindent\textbf{Seeding strategy: } Various seeding strategies attempt to optimize different goals such as aesthetics,
clutter reduction, capturing flow features, etc \cite{sane20}. In this paper, the primary consideration for pathline seeding take into account where
one can have the highest return in informational value regarding the water
movement. Because most of our training data and random beach images that one can find on the web are taken in landscape mode, seeds are distributed uniformly along the horizontal axis. Conversely, the water body is usually in the middle, with the sky on top and the beach on the bottom halves
of the framing. We therefore use a Gaussian distribution of seeds along the vertical axis. This is a form of importance sampling \cite{burger2008importance}. During the training process, we use importance seeding to maximize pathlines that track water movement, and reduce irrelevant information such as the beach or cloud movement in the sky. During the testing process, we use regular sampling since there is no guarantee that the test data will also be in landscape mode.

\begin{figure}[tb] 
    \centering
  \subfloat[Regular grid seeding]{%
      \includegraphics[width=0.49\linewidth]{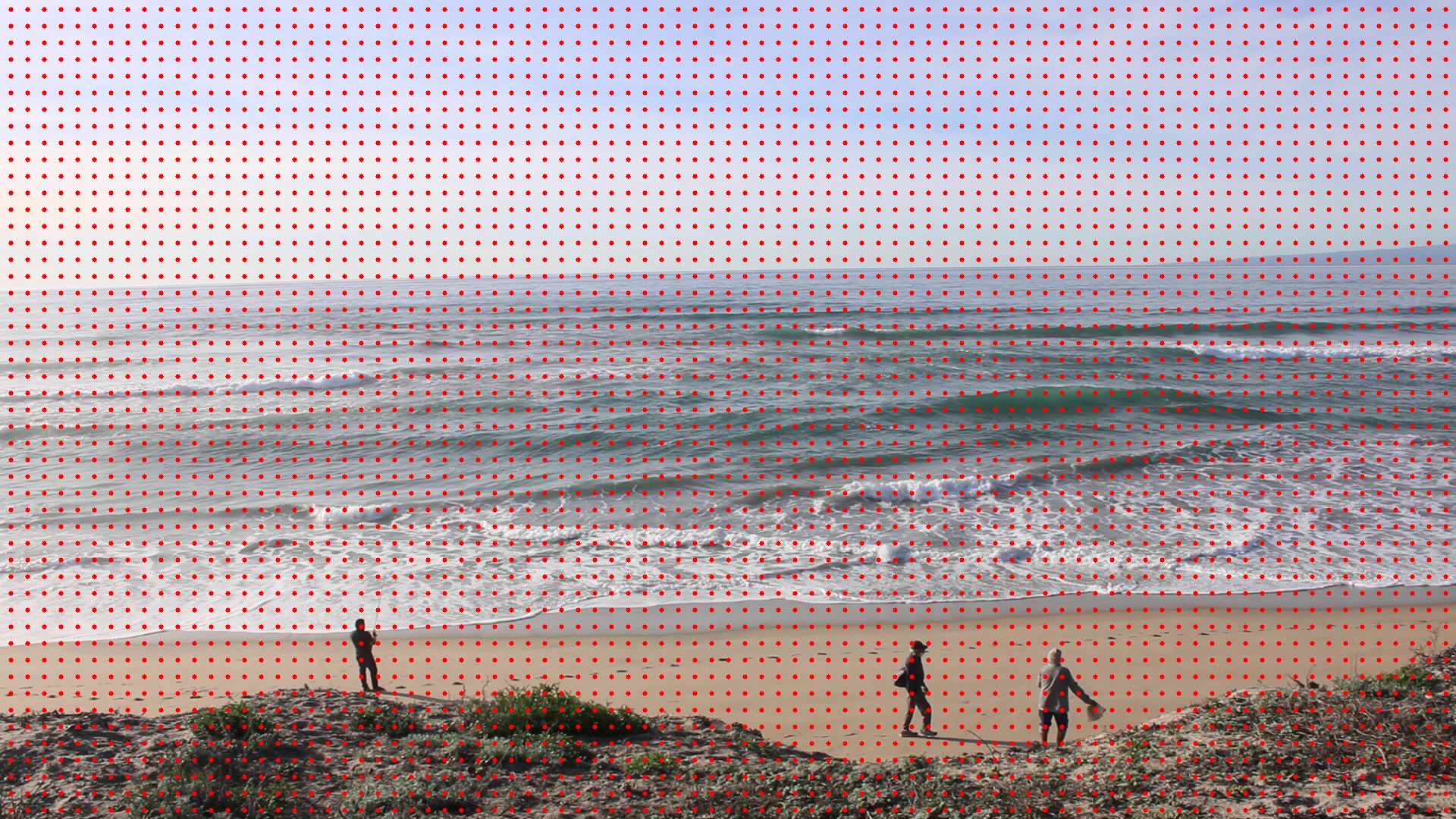}}
    \hfill
%   \subfloat[Random seeding]{%
%         \includegraphics[width=0.49\linewidth]{figures/seed_points/seed_points_random.png}}\\
%   \subfloat[2D normal seeding]{%
%         \includegraphics[width=0.49\linewidth]{figures/seed_points/seed_points_2D_gaussian.png}}
%     \hfill
  \subfloat[Importance seeding]{%
        \includegraphics[width=0.49\linewidth]{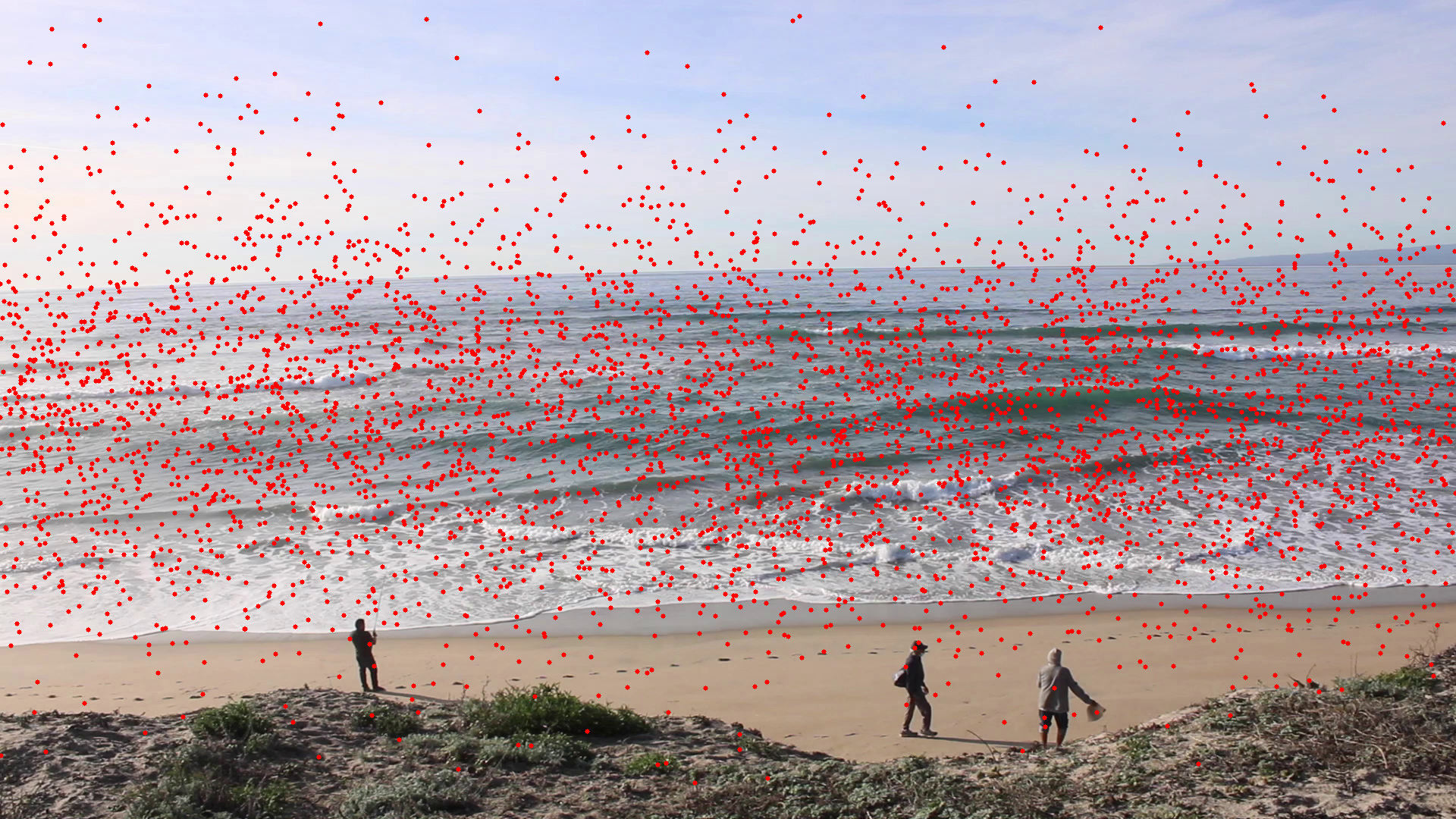}}
  \caption{\textbf{Seeding strategies:} We used regular grid to seed pathlines when testing. For training we used importance seeding.}
  \label{seeding_strategies} 
\end{figure}

% \vspace{2pt}\noindent\textbf{Amount of training data: }

% We also noticed that although the rip current was highlighted by the colored seed points, the points belonged to multiple clusters as shown in \ref{fig:seed_point_projection}. Once we selected the cluster on the rip we noticed that there were more false positives than RipViz appraoch. 

% We noticed that the clusters in the scatter plot were not separated enough [?]. More False positives. 
% \begin{sidewaysfigure*}
\begin{figure*}
    \centering
    \includegraphics[scale=0.078]{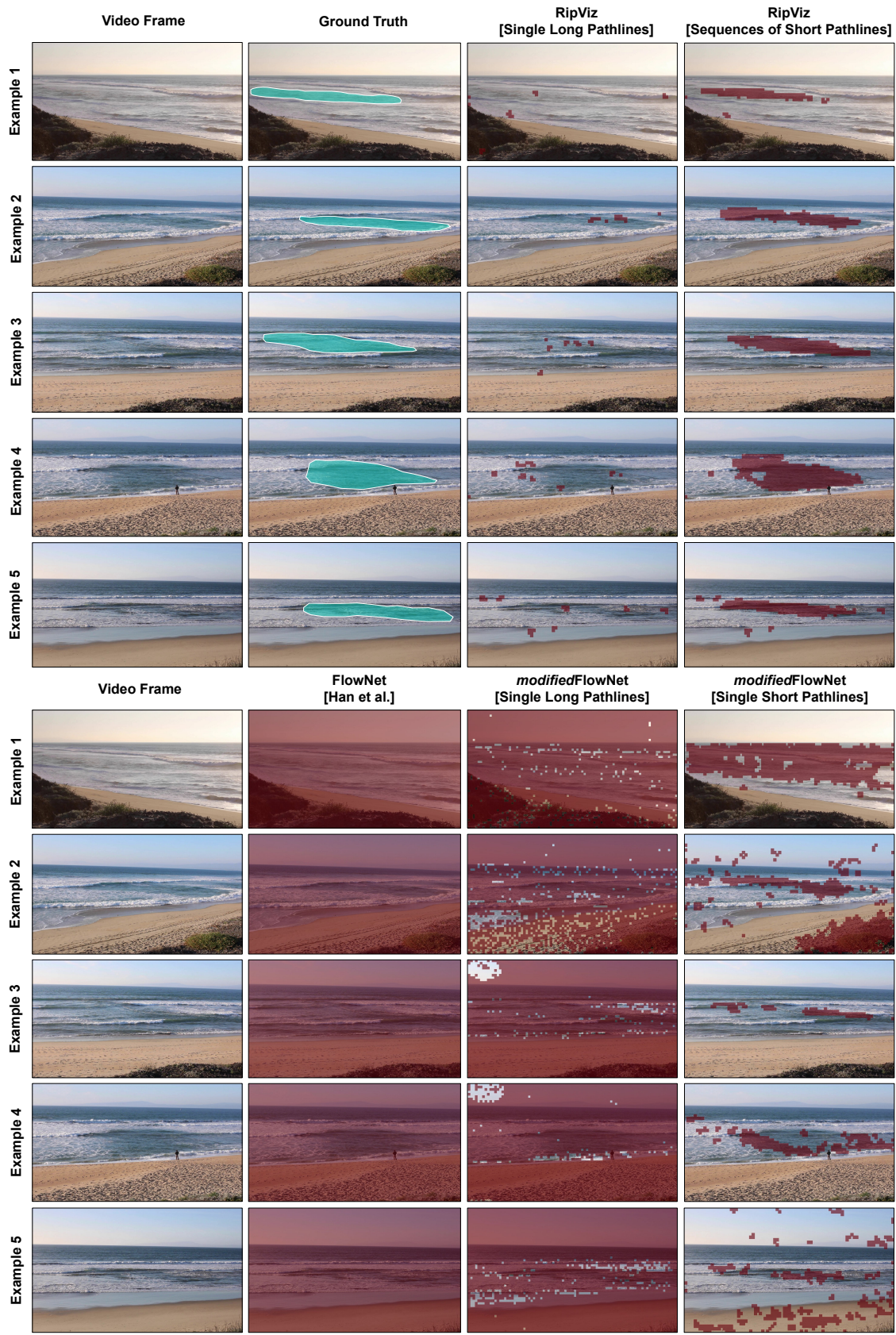}
    \caption{\textbf{Comparison with FlowNet and different pathline parameters:} The \textit{Video Frame} column shows a representative frame of the video clip while the \textit{Ground Truth} column shows the expert's ground truth estimate. Notice that FlowNet without modifications (\textit{FlowNet} Column) incorrectly clusters all seed points into a single group. Both \textit{modified} Flownet and RipViz when trained with longer pathlines were not able to detect the rip current. When \textit{modified} FlowNet was trained on single shorter pathlines ($n = 900$), we were able to find a cluster representing the rip current but not as definitive. RipViz trained with shorter pathline sequences was able to detect the rip currents more precisely. Best viewed in color.}
    \label{fig:flownet_comparison}
\end{figure*}

\subsection{Comparison with FlowNet \added[id=A1]{using Single Long and Sequences of Short Pathlines}\label{s:flowNet_comparison}}

We compared RipViz with FlowNet \cite{han2018flownet}, a deep learning based streamline clustering method. In FlowNet, $3D$ streamline features were learned by an autoencoder. \added[id=A-2]{These 3D streamlines were traced across the full extent of the data.} \added[id=A-2]{Additionally, they traced one streamline per seed point.} The trained autoencoder was used to generate 1D representations of 3D streamlines. The dimensionality of these $1D$ representations is further reduced by t-distributed stochastic neighbor embedding (t-SNE) \cite{tsne}. The resulting low dimensional vectors are then clustered by using DBSCAN \cite{DBSCAN_ester1996density}. An interactive user interface was used to find the desired clustering by changing the maximum distance between two feature descriptors ($eps$) and the minimum number of samples in each cluster ($ms$).The autoencoder in FlowNet is trained with a unweighted binary cross-entropy function (i.e., $w_0 = w_1 = 1$). 

\added[id=A-2]{In order to compare with FlowNet, we transformed our long $2D$ pathlines to $3D$ by adding a time axis as the $z$ axis \cite{theisel2003feature}. However, since our pathlines are sparsely distributed in the $3D$ binary volume, we found that the FlowNet model with the unweighted binary cross entropy did not learn to differentiate the pathlines. Resulting in a single cluster for all the pathlines.  We show the clustering result in Figure \ref{fig:flownet_comparison}. Additionally, the training time was relatively long. } \deleted[id=A-2]{Furthermore, FlowNet only learns the spatial features of the streamlines. In contrast, our method uses $2D$ pathlines and a weighted binary cross-entropy function and also learns the spatiotemporal features of pathlines by utilizing LSTM layers.} A proper comparison requires modification of FlowNet to work with our application domain. The modifications are discussed in Appendix A. 

\added[id=A-2]{We trained the modified FlowNet model with long pathlines, integrated across the entirety of the video, and with single short pathlines ($n = 900$).} The output of the \textit{modified} FlowNet is shown in Figure \ref{fig:flownet_comparison}.
The first column shows a representative frame of each video. \deleted[id=A-2]{The \textit{t-SNE view} column of Figure \ref{fig:flownet_comparison} shows the clusters, color coded according to cluster membership, in the t-SNE feature space. The \textit{projection} column of the same figure shows those clusters projected back on to the video frame. To do this projection we mapped the seed point of each pathline of the associated cluster back on to the frame using the same color coding. The column \textit{Rip cluster selection} shows the projection of cluster/clusters that are only on the rip current. The fourth column, \textit{RipViz}, shows the output of our method.} 
\added[id=A1]{We see that without modifications, FlowNet lumps all the pathlines into one cluster.  Both the modified FlowNet and RipViz, when fed with single pathlines that run the entirety of the video, also fail to identify the rips.  When fed with shorter pathlines ($n = 900$), modified FlowNet start to show signs of rip currents but includes significant FPs as well.  Only, when sequence of pathlines as used in RipViz do we see the rip currents clearly.} \deleted[id=A-2]{The hyper-parameters used in the clustering step of FlowNet by the DBSCAN algorithm, $eps$ and $ms$, are shown in Table. } 

\deleted[id=A1]{We noticed that in most instances, \textit{modified} FlowNet was able to highlight the rip region as shown in Figure \ref{fig:flownet_comparison}. However, the number of FPs outside the rip region were also significantly high. This is evident in Figure \ref{fig:flownet_comparison}, examples 1 and 3. To compare with RipViz, }
\added[id=A1]{In order to obtain a quantitative comparison between \textit{modified} FlowNet with shorter pathlines and RipViz, we calculated their $F_1$ scores.}
\deleted[id=A1]{from the cluster/s on the rip current from \textit{modified} FlowNet as shown in the \textit{Rip cluster selection} column of Figure \ref{fig:flownet_comparison}. We used the $F_1$ score metric used in equation \ref{eq:f_1_score}.}
If a seed point is flagged as as a rip by either method is within the expert annotated rip current boundary then we mark it as TP, otherwise we mark it as FP. If seed points within the rip are not selected then we mark those points FN.
%\blue{FN and FP may be switched here. double checkk}
We found that the $F_1$ score for \textit{modified} FlowNet was $0.32$ compared to the $F_1$ score of $0.85$ for RipViz  as shown in Table \ref{table:results_summary}. We hypothesize the low $F_1$ score for the \textit{modified} FlowNet was due to its high FP rate.

\deleted[id=A-2]{We also noticed that color coded seed points on the rip current can sometimes belong to multiple clusters. In Figure \ref{fig:flownet_comparison} example 2, \textit{rip cluster selection} column, the seed points on the rip current belongs to two different clusters, color-coded in red and pink. As expected in the \textit{t-SNE view} column for example 2, the two clusters were spatially close. This may introduce some ambiguity to the viewer. In contrast, RipViz tries to identify anomalous pathlines without a clustering step.}

% are the Although the \textit{modified} FlowNet highlighted the rip region, we noticed that the pathlines on the rip current could in some instances belong to multiple clusters. In Figure \ref{fig:flownet_comparison} example 2, rip cluster selection column, we noticed that the pathlines on the rip current consist of two different clusters, color-coded in red and pink. As expected in the t-SNE view, the two clusters were spatially close. This may introduce some ambiguity to the viewer. In contrast, RipViz tries to identify anomalous pathlines without a clustering step. 

Furthermore, we found tuning the two hyperparamters $eps$ and $ms$ for \added[id=A-2]{the} clustering step was \deleted[id=A-2]{a} crucial \deleted[id=A-2]{step} in finding the appropriate clusters. We noticed having some domain knowledge was helpful for the user when exploring these two hyperparameters to find the ideal clustering. \deleted[id=A-2]{As shown in  Table  the two hyperparameters differed for each video example in Figure \ref{fig:flownet_comparison}.} In contrast, RipViz is automated and does not require any \added[id=A-2]{user input} \deleted[id=A-2]{hyperparameter tuning} during run time. \deleted[id=A-2]{as shown in Table \ref{table:results_summary}}

Both \textit{modified} FlowNet and RipViz are based on autoencoders. However, RipViz learns how each pathline of the same seed point was related in time by learning spatiotemporal features of pathline sequences. This allows RipViz to learn ocean flow behavior more effectively. On the other hand, \textit{modified} FlowNet does not learn how pathlines are related in time. Therefore, we hypothesize that using pathline sequences is better for learning quasi-periodic behavior, such as in ocean flow.

% Although the \textit{modified} FlowNet highlights the rip region we noticed that the pathlines can some instances belong to multiple clusters. 
% RipViz only outputs the pathlines that it labels as rip currents. 

% \begin{table}[]
%     \centering
%     \begin{tabular}{|c | c c|} 
%          \hline
%           & $eps$ & $ms$  \\ %[0.5ex] 
%          \hline\hline
%          Example 1 & 3.65 & 93 \\ 
%          \hline
%          Example 2 & 3.75 & 93 \\ 
%          \hline
%          Example 3 & 3.94 & 81 \\ 
%          \hline
%          Example 4 & 3.70 & 91 \\ 
%          \hline
%          Example 5 & 3.30 & 83 \\ 
%          \hline
%     \end{tabular}
%     \caption{Hyper-parameters for \textit{modified} FlowNet. The video frame name refers to the videos shown in Figure \ref{fig:flownet_comparison}. Notice that for each example hyper parameters were different. \added[id=A-2]{WILL MOVE TO SUPP MATERIALS: SEEMS NOT RELEVANT TO THE PAPER} }
%     \label{table:eps_and_ms}
% \end{table}

\subsection{Comparison with existing rip detection methods \label{s:rip_detection_comparison}}

We compared RipViz with existing rip current detection methods as shown in Figure \ref{fig:previous_methods_comparison}. The first \added[id=A-2]{and second} columns of Figure \ref{fig:previous_methods_comparison} \added[id=A-2]{show }a representative frame of the video\added[id=A-2]{ and the expert drawn ground truth}. The remaining columns show the output of the object detector \cite{de2021automated}, RipViz,
filtered arrow glyphs \cite{mori2022flow}, 
filtered color maps \cite{mori2022flow}, and
timelines \cite{mori2022flow}, respectively.

% There are several existing methods used to detect rip currents. 

\subsubsection{Comparison with behavior based methods}

% \begin{figure}[tb]
%     \centering
%   \subfloat[Good timeline placement\label{good_timeline}]{%
%       \includegraphics[width=0.49\linewidth]{figures/timeline_comparison/MVI_9066_start_(390,543)_end_(1357,426)_seed_pts_2000_interval_30s.00_02_56_10.Still002.png}}
%     \hfill
%   \subfloat[Poor timeline placement\label{bad_timeline}]{%
%         \includegraphics[width=0.49\linewidth]{figures/timeline_comparison/MVI_9066_start_(390,543)_end_(1357,426)_seed_pts_2000_interval_30s.00_02_56_10.Still001.png}}
%   \caption{This figure compares good and poor timeline placements in the same test video. The grey and blue lines indicate the initial and final positions of the timeline respectively. In Figure \ref{good_timeline} the timeline indicates the presence of the rip. However, in Figure \ref{bad_timeline} the timeline fails to detect the rip. \added[id=A-2]{WILL MOVE TO SUPP MATERIALS. SEEMS NOT RELEVANT TO THE PAPER.}}
%   \label{fig1} 
% \end{figure}

% \begin{figure}[tb]
%     \centering
%   \subfloat[Optimal timeline placement\label{good_timeline}]{%
%       \includegraphics[draft,width=0.49\linewidth]{long_rip.png}}
%     \hfill
%   \subfloat[Non-optimal timeline placement\label{bad_timeline}]{%
%         \includegraphics[draft,width=0.49\linewidth]{long-rip.png}}
%   \caption{Comparison between object detector and RipViz on long rip current}
%   \label{fig:long_rip} 
% \end{figure}

\begin{figure*}
    \centering
    \includegraphics[scale=0.078]{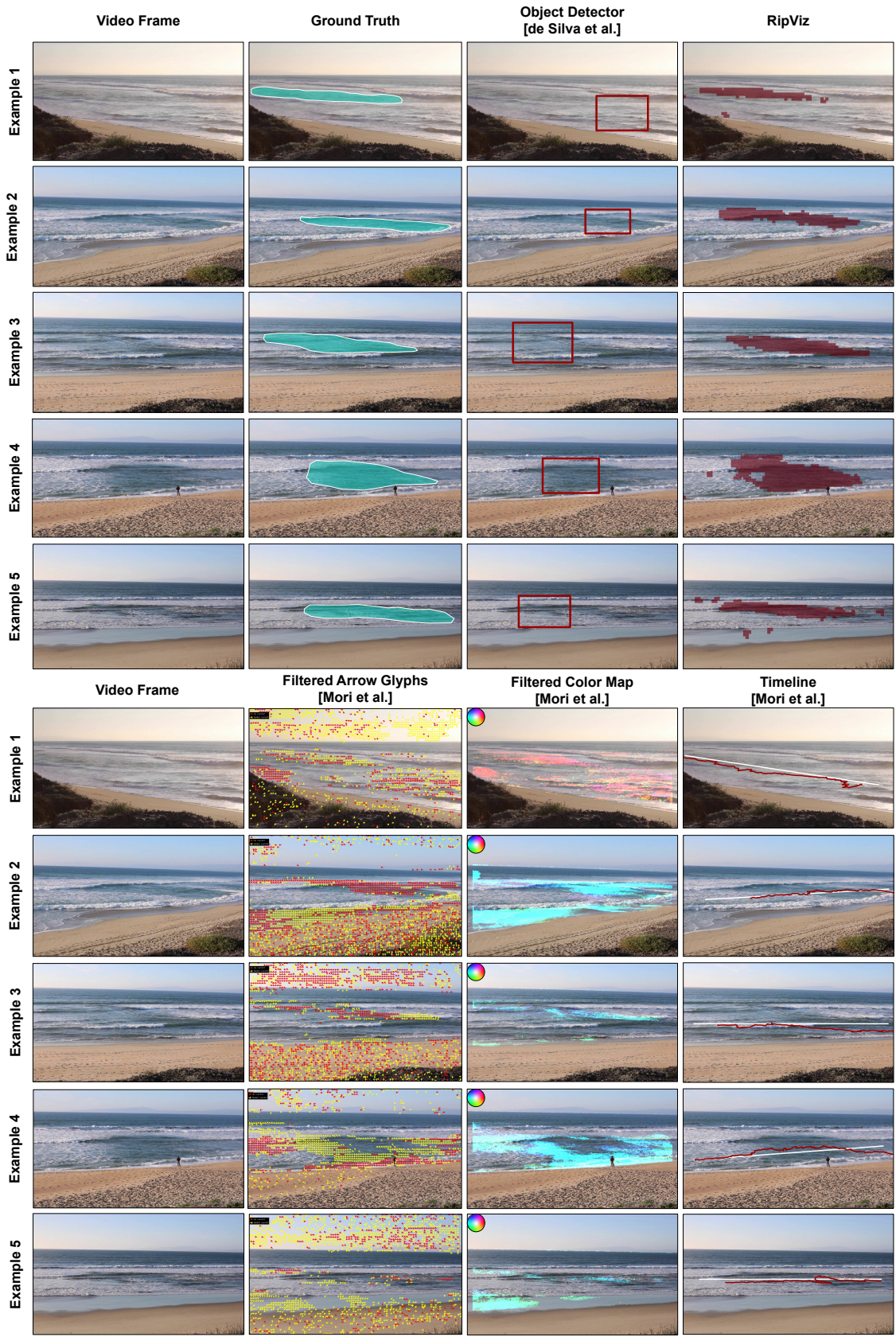}
    \caption{\textbf{Comparison with prior rip current detection methods:} The \textit{Video Frame} and \textit{Ground Truth}  columns show a representative frame of the video clip and the expert ground truth estimate respectively. The remaining columns show object detector, timeline, colormap, arrow glyph, and RipViz, respectively. Note that the timeline %filtered color map, and arrow glyph methods 
    method requires user
    %\blue{example 5, timeline, why is it oriented that way and not parallel to beach?} 
    input to specify its initial placement, and that Filtered Arrow Glyphs did not filter out erroneous glyphs from the sky, beach, and non-rip area of the water. Best viewed in color.}
    
    \label{fig:previous_methods_comparison}
\end{figure*}

\vspace{2pt}\noindent\textbf{Timelines: } Mori et al. \cite{mori2022flow}  proposed to use timelines to detect and visualize rip currents. They placed timelines parallel to the beach and traced the points on the timeline using optical flow to update its position. They observed the shape of the timeline as it gets dragged by the rip current. Timelines get deformed and extend to the rip channel above its initial position, as shown in Figure \ref{fig:previous_methods_comparison}. 

However, the initial placement of the timeline is a major contributing factor for timelines to be successful. 
% If the timeline was placed in a good initial position by a domain expert, visualization could indicate the presence of the rip current, as shown in Figure \ref{good_timeline}. However, if the timeline was placed poorly, as shown in Figure \ref{bad_timeline}, we could not visualize the rip current. 
All the timelines indicated in Figure \ref{fig:previous_methods_comparison} are placed optimally.
%We found the optimal position of the timeline by using domain expertise and sometimes by trial and error. 
To compare the timeline method with RipViz, we calculated the $F_1$ score. If the timeline was able to visualize the rip current in a video, then we counted it as TP, otherwise it is counted as FN. We found that properly placed timelines can visualize rip currents in most cases. However, the user has to specify the good initial placement for each video. In contrast, RipViz is automated, and no user input was needed during run time, as shown in Table \ref{table:results_summary}.
% \red{add stats} \red{Also need to add reasoning as to why timelines are failing even when the rip current is present.}

\vspace{2pt}\noindent\textbf{Direction based clustering: }
Philip and Pang \cite{philip2016detecting}
%obtained an unsteady flow field from the video and
hypothesized that the rip current is directly opposite the dominant flow in the vector field, which corresponds to the incoming wave direction.
%They grouped vectors based on the direction and magnitude to visualize the rip current. 
Places where the vectors were opposing the majority flow are potential rip zones.  Areas with sufficiently large clusters of such vectors and large enough magnitude were highlighted as rip zones. 

Mori et al. \cite{mori2022flow} used a similar approach and mapped direction to color and magnitude to hue for better visualization. As shown Filtered Color Map column of Figure \ref{fig:previous_methods_comparison}, the method can detect the location of the rip current. However, this method also highlighted the swash zone, the shallow part of the beach, as evident in examples 1-3. 
%Furthermore, the filtered color map method works best when performed only on the ocean parts of the video. The ocean parts are user defined by a binary mask. In contrast, for RipViz there is no such pre-processing. 
We found that while the rip current was highlighted in most cases, the number of false positive pixels tend to be high. In order to compare with RipViz, we calculated the $F_1$ score. As shown in Table \ref{table:results_summary} filtered color map resulted in a $F_1$ score of 0.28, due to its high false positive rate. 

Mori et al. \cite{mori2022flow} also proposed arrow glyphs  to visualize rip currents. The output of this method is shown in Arrow Glyph column of Figure \ref{fig:previous_methods_comparison}. We noticed that the arrow glyph method is more susceptible to noise in the flow field compared to other methods and RipViz. This is evident by the arrows projected on the sky and the beach although there is a little movement on those parts of the video. Although arrow glyph method highlighted the rip current in the majority of the videos, there were also a large number of false positive detections. Similarly we calculated the $F_1$ score.
As indicated in Table \ref{table:results_summary}, filtered arrow glyph had a $F_1$ score of 0.16 due is relatively high false positive rate.

\subsubsection{Comparison with appearance based methods}

% \begin{figure}[tb]
%     \centering
%   \subfloat[Frame\label{1a}]{%
%       \includegraphics[width=0.49\linewidth]{figures/timex_comparison/3000.jpg}}
%     \hfill
%   \subfloat[Timex\label{1b}]{%
%         \includegraphics[width=0.49\linewidth]{figures/timex_comparison/timex_MVI_9071.png}}
%   \caption{This figure compares timex images.}
%   \label{fig1} 
% \end{figure}

\vspace{2pt}\noindent\textbf{Object Detectors: }
de Silva et al. \cite{de2021automated} used object detectors to detect rip currents. They trained a deep learning based object detector, Faster R-CNN \cite{ren2015faster}, with images of rip currents. The images they used had a clear visual signature for bathymetry controlled rip currents with a darker region between breaking waves. Then they used the model to predict the location of rip currents, by overlaying a bounding box on the rip currents on videos. The output of the method is shown in the \textit{Object Detector} column of Figure \ref{fig:previous_methods_comparison}. 

We noticed that the object detector can detect parts of rip currents as shown in examples 2-5 in Figure \ref{fig:previous_methods_comparison}.
%if the appearance of the rip current is similar to the training data. 
However, rip currents have different appearances \cite{castelle2016rip}.  As evident in example 1, 
%if the appearance is slightly different from the training data, the object detectors failed to detect the rip current.
it is possible for an objector detector to miss the actual rip current location.
% However, different types of rips have different appearance \cite{castelle} XXX.
% The model in \cite{de2021automated} was trained on rips known as bathymetry controlled rips only.  Hence it won't detect other types of rips using their appearance only.  Furthermore, even for bathymetry controlled rips, if the appearance of the rip current is not obvious, then object detectors could failed. In example 2 of Figure \ref{fig:previous_methods_comparison}, the object detector failed to detect the rip current.
In contrast, RipViz is trained on pathlines that capture the {\em behavior} rather than appearance of rip currents.  Hence, it was able to detect the rip current in all the examples.

Furthermore, we noticed that bounding boxes only detects the general area of the rip current.
\added[id=A1]{Sometimes the bounding boxes can cover parts of the beach or miss parts of the rip especially with long elongated rips.}
\deleted[id=A1]{Sometimes, as shown in example 5 of Figure \ref{fig:previous_methods_comparison}, the bounding boxes can cover part of the beach as well. Also, bounding boxes in some instances detect only part of the rip current, particularly when the rip current is long as shown in examples 1 and 3 of Figure \ref{fig:previous_methods_comparison}.}
We also noticed that the information such as the curvature of the rip current cannot be gleaned from bounding boxes alone.
\deleted[id=A1]{as shown in Figure \ref{fig:previous_methods_comparison}.}
In order to compare with RipViz, we calculated the $F_1$ score. We counted how many pixels were covered by the bounding box. Out of those, rip pixels are counted as TP, the remaining pixels were counted as FP. If the object detector does not predict a bounding box, then we counted the pixels within the boundary as FN. 
%\blue{i think you mean FN and not FP here.} 
As shown in Table \ref{table:results_summary}, the bounding boxes resulted in a $F_1$ score of $0.43$ compared to $0.85$ for RipViz. We attribute this lower $F_1$ score to the large number of false positives and false negatives generated respectively by the non-rip areas of the bounding box and by part of the rip current not being covered by the bounding box.

\begin{figure}[t]
    \centering
    \includegraphics[width=1\linewidth]{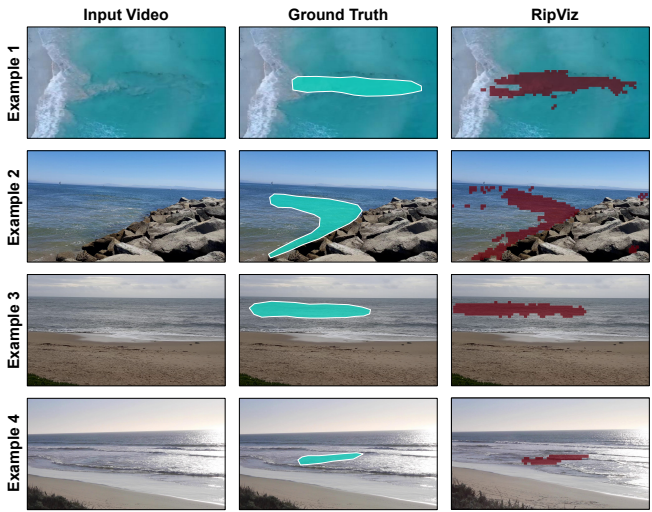}
    \caption{\textbf{RipViz on novel rip current types:} Examples 1 and 2 are sediment rip and structural rip, respectively. Note that they have very different visual characteristics compared to bathymetry rips. No published object detectors exist for these two types of rips. However, notice that RipViz could detect rip currents regardless of their varying appearance. Examples 3 and 4 illustrate two rip currents where object detectors failed due to a lack of expected visual features. In these two examples, RipViz can still detect the rip current because it uses \textit{behavior} rather than appearance to detect rip currents. Best viewed in color.}
    \label{fig:other_types_of_rips}
\end{figure}

\added[id=A-2]{Additionally, we tested RipViz on other types of rip currents where currently there are no published object detector available. In Figure \ref{fig:other_types_of_rips}, we show a few examples of such rip current types. Experts categorize the rip currents in the first and second row as sediment and structural rips. However, as shown in Figure \ref{fig:other_types_of_rips} RipViz could detect these rip currents regardless of their appearance. More importantly, no additional training data were needed for sediment and structural rips. Examples 3 and 4 illustrate two rip currents where object detectors failed due to a lack of expected visual features. In these two examples, RipViz detected the rip currents because it uses \textit{behavior} rather than appearance to detect rip currents.}

\deleted[id=A-2]{\vspace{2pt}\noindent\textbf{Timex images: } }
\deleted[id=A-2]{
%Say that timex images were used by the coastal community for years to visualize rip currents. 
In order to interpret the time averaged images, a human expert was needed. Maryan et al.  proposed to overcome the need for a human expert by using ML. Similar to the experiments in  and , we also found that the bounding boxes from  did not detect rip currents in most of our data. The images that was used to train their model were grey scale images that was more suited to predict on similar type of remote sensing data. We hypothesize that this domain difference was the primary reason for their method not detecting the rip currents in our data. }
%domain difference between the training data used in \cite{maryan2019machine} and our test data is a contributing factor for its low performance. 
% \subsection{Run time analysis if we have time \label{s:run_time_analysis}}
% Need to show a graph of how long it takes to make a prediction. Only do this if we have time. Most of our experiments were conducted on a Tesla V100-SXM2-16GB graphical processing unit. 
% \blue{if timex of your recent dataset doesn't show the signatures of rips, it may be because the rips are weaker than those in their paper i.e. waves occasionally break over the rip channel in our video; plus the rips were not stationary or always present throughout the video.  probably not because of grayscale.. i mean you could turn the timex to grayscale before testing on ml, but i doubt that's the reason}
\begin{table}[]

    \renewcommand{\arraystretch}{1.5}
    \centering
    \begin{tabular}{|c |  c c|} 
         \hline
         Method & \begin{turn}{90} Automated?\end{turn}  & \begin{turn}{90} F$_1$ Score\end{turn}   \\ %[0.5ex] 
         \hline\hline
         Object Detector \cite{de2021automated}  & \textbf{\textit{yes}}  & 0.43\\ 
        %  \hline
        %  Maryan et al. \cite{maryan2019machine}  & \textbf{\textit{yes}}  & 0.15\\ 
         \hline
         Timelines \cite{mori2022flow}  & \textit{no} & $0.72$ \\ 
         \hline
         Filtered color map \cite{mori2022flow}  & \textbf{\textit{yes}}  & 0.28  \\ 
         \hline
         Filtered arrow glyph \cite{mori2022flow}  & \textbf{\textit{yes}}  & 0.16 \\ 
         \hline
          FlowNet\cite{han2018flownet}  & \textit{no}  & 0.16 \\
          \hline
         \textit{modified} FlowNet [single long pathlines]  & \textit{no}  & 0.19 \\ 
         \hline
         \textit{modified} FlowNet [single short pathlines]  & \textit{no}   &  0.32 \\ 
         \hline
         RipViz [single long pathlines]  & \textbf{\textit{yes}}  & 0.08 \\     
         \hline
          RipViz [sequence of short pathlines] & \textbf{\textit{yes}} & \textbf{0.85}  \\ 
         \hline
    \end{tabular}
    \caption{\textbf{Results summary:} Notice that RipViz has the highest $F_1$ score, and does not rely on user input at run time.}
    \label{table:results_summary}
\end{table}

\subsection{\added[id=A-2]{Discoveries made by Experts using RipViz} \label{s:discovery}}

% how rip current boundaries are usually estimated. 

\added[id=A-2]{Rip current researchers use their expert knowledge to identify the boundary of rip currents by observing the appearance of visual features such as gaps in breaking waves or sediment plumes. However, sometimes parts of the rip current lack these visual features, making it challenging for the expert to identify the entire extent of the rip current. } 

% how ripviz helps to identify the rip current boundary. 

\begin{figure*}[tb]
    \centering
    \includegraphics[scale=0.078]{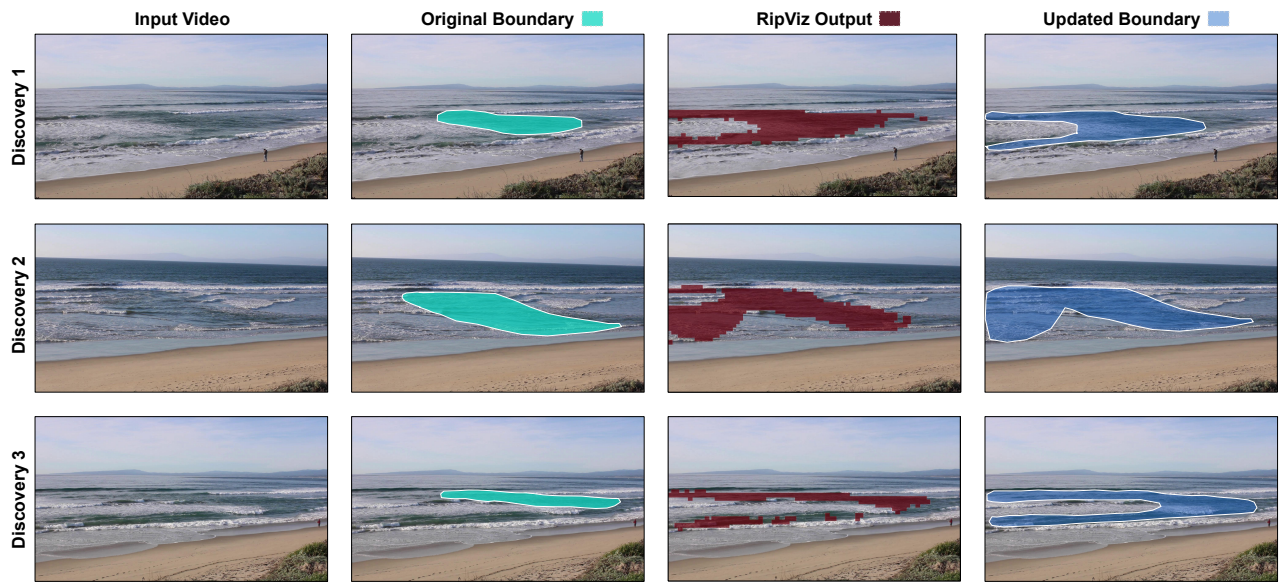}
    \caption{\textbf{Discoveries made by Experts using RipViz:} Experts estimated the rip current boundary (\textit{Original Boundary }Column) by observing only the input video (\textit{Input Video} Column). After examining the output of RipViz (\textit{RipViz Output} column), the experts updated their original rip current boundary estimate to include parts of the rip current that are not readily observable in the input video due to a lack of distinct visual features (\textit{Updated Boundary} column). In the supplementary materials, we included a complete table of discoveries made on our test data.Best viewed in color.  }
    \label{fig:discovery}
\end{figure*}

\added[id=A-2]{In contrast to appearance-based identification, RipViz uses the behavior of rip currents, not appearance, to detect rip currents. Experts can use RipViz as a visualization tool to determine the entire boundary of the rip current when visual features are lacking. In Figure \ref{fig:discovery}, we demonstrate a few use cases where the rip current expert's original boundary estimate was updated after examining the visualization provided by RipViz. \added[id=A1]{The updated boundaries now include feeder currents near shore and what appears to be a circulating pattern farther offshore in Discovery 1 and Discovery 3, as well as a weaker neighboring rip that merged with the dominant rip in Discovery 2.  Incidentally, Discovery 1 and 3 both show circulating rips where the guidance from beach signage to swim parallel to shore may not always work \cite{macmahan11}. } We offer the complete table of discoveries made on our test data in the supplementary materials. } 

\subsection{\added[id=A-2]{User Study} \label{s:user_study}}
\added[id=A-2]{We conducted a user study to better understand how non-experts can use RipViz as a tool to become more aware of rip currents. We grouped $400$ non-experts into two groups; one group was trained only with beach warning signs of rip currents, the other group was trained with RipViz. We then showed each participant a randomly picked video of a rip current and asked them if there was a rip current present. In the group trained with beach signs, $24\%$ failed to recognize the existence of the rip current in the video. In the group trained with RipViz, only $14\%$ were unable to recognize the presence of the rip current in the video.}

\added[id=A-2]{We performed a second experiment to better understand if the participants could locate the rip current within the video. Both groups were given multiple choices of rip current boundary estimates and were asked to pick the correct one. In the group trained  with beach signs, only $34\%$ could choose the correct boundary. In the group trained with RipViz, $78\%$ could select the correct boundary. The non-experts were acquired using Mechanical Turk \cite{mturk_1, mturk_2}, with basic screening for reliable workers, and paid \$0.10-\$0.20 per task.}

% We conducted a user study using mturk

%To measure human accuracy of identifying rip currents, annotators were asked to draw bounding boxes around places they believe to have rip currents. We sampled  every tenth frame from our video test set and randomized presentation order across all positive and negative examples. Human annotators were not carefully trained, instead they were provided three positive and three negative examples, roughly the amount of information which might fit on a sign at the beach. Annotators were acquired using Mechanical Turk, an online market where jobs are posted for workers \cite{mturk_1, mturk_2}, with basic screening for reliable workers, and paid \$0.10 per image. 

\subsection{\added[id=A-2]{Sensitivity to video framing} \label{s:user_study}}
\begin{figure*}
    \centering
    \includegraphics[scale=0.07]{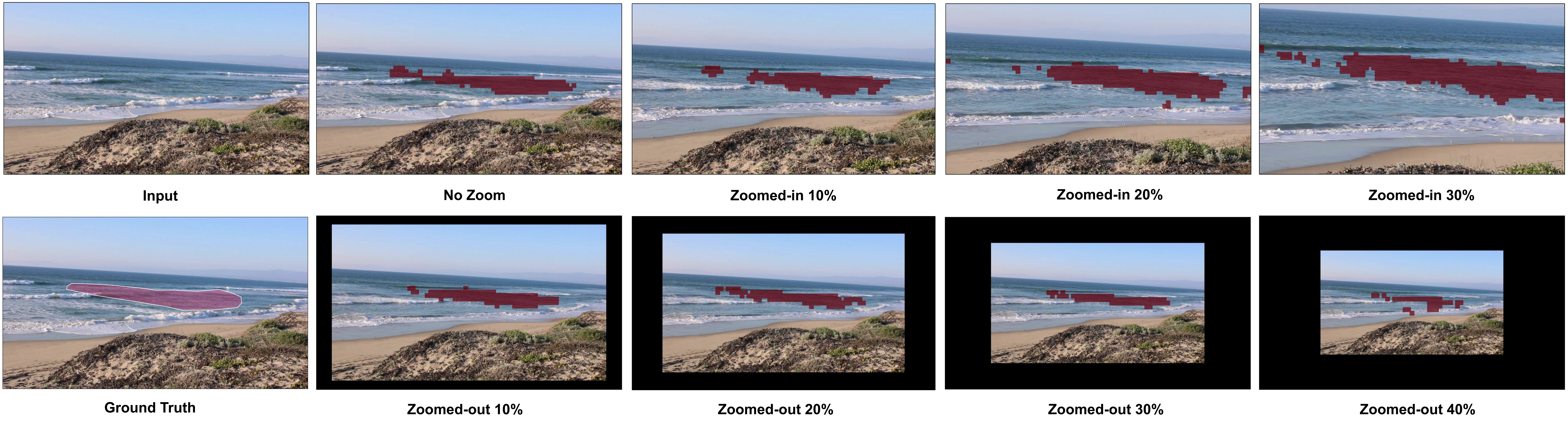}
    \caption{\textbf{RipViz is less sensitive to different video framing:} We changed the scale of the rip current feature by zooming in and out from the original video, while keeping the aspect ratio of the video unchanged. Notice that for instances where the rip current feature is larger or smaller than the training data (within reasonable bounds), RipViz was still able to detect the rip current feature.  }
    \label{fig:zoom_image}
\end{figure*}

\added[id=A-2]{Some of the parameters are dependent on the camera framing which would include distance of camera to the water and focal length or zoom factor of the lens. Alternatively, we can think about the width of the beach that is visible in the frame. We base our framing by examining a number of existing surf webcams (e.g. webcoos.com) and planned webcam installations. We experimented on how sensitive our parameters are to changes in camera framing. In the Figure \ref{fig:zoom_image}, we show effects of varying the framing while keeping our current set of parameters. Recall that during testing, we trace the same number of seed points that are distributed in a regular grid.  Based on the results shown in Figure \ref{fig:zoom_image}, we believe the parameters are still valid within 20-30\% change in camera framing. We found the $F_1$ score to be $0.82\pm0.3$ for all the examples, without much deviation.}

\section{Summary and Remarks \label{s:conclusions_and_future_work}}

\added[id=A1]{
As Ben Schneiderman succinctly captured in his quote:
``The purpose of visualization is insight, not pictures'', our goal is to make apparent what may not be visible to the untrained eyes.  The focus of this work is to first find the feature of interest (rip) and then present them in a simple, easy to understand manner by highlighting their location directly on the video. 
This capabality is encapsulated in}
\deleted[id=A1]{We presented} RipViz, a hybrid feature detection method that combined ML and flow analysis to extract rip currents from stationary videos. We used shorter pathline sequences to capture the flow behavior in a  noisy \added[id=A1]{quasi-periodic }flow field. Then we used an LSTM autoencoder to learn the behavior of normal ocean pathline sequences. The trained model allowed us to label pathline sequences belonging to rip currents as anomalous by comparing the reconstruction error. \added[id=A1]{By framing the rip detection problem as an anomalous flow detection problem, the onerous task of finding and labeling training datasets for each type of rip current is also greatly reduced}. 

The Visualization community has used deep learning methods to learn flow behavior from pathlines. In particular, existing literature uses autoencoders to learn from flow data. The authors use single long pathlines/streamlines per seed point as input to their deep learning models. However, straight-forward learning
based on traditional pathlines did not work for our application due to the quasi-periodic flow fields such
as those found near the surf zone. Therefore, we adapted and innovated on the existing deep learning
methods to use a sequence of pathlines per seed point instead\\

\noindent The main contribution of this paper is: 
\begin{itemize}
    \item A hybrid feature detection method that combines machine learning and flow analysis techniques to automatically find and visualize dangerous rip currents. 
\end{itemize}
  
\added[id=A1]{In order to realize this, the following innovations are necessary:}
\begin{itemize}
        \item \added[id=A-2]{For flow fields with a quasi-periodic behavior, such as ocean flow, working with a {\em sequence} of shorter pathlines is better than a single long  pathline. } 
        
        \item \added[id=A-2]{A weighted binary cross entropy, unlike the unweighted binary cross entropy used in previous works, is more effective when learning from sparsely distributed pathlines generated from ocean scenes.  }
\end{itemize}

% \deleted[id=A-2]{
% Aside from RipViz itself,
% the contributions from this work include:}
% \begin{enumerate}
%     \item \deleted[id=A-2]{When working with real world datasets, noise is a concern.  Using shorter pathlines to learn about noisy flow fields is better than longer pathlines.This is discussed further in Appendix B.}
%     \item \deleted[id=A-2]{For certain flow fields, e.g. with quasi-periodic behavior, working with pathline sequences is better than single pathlines} 
%     %(or splitting a single pathline into multiple pieces).
%   \deleted[id=A-2]{ The pathline sequences allows the LSTM based autoencoder to learn about flow behavior more effectively.}
%     \item \deleted[id=A-2]{We empirically determined the weights for the loss function for our driving application}
%     \deleted[id=A-2]{and showed how such a weighted binary cross-entropy function can better learn from sparsely distributed pathlines.}
%     \item \deleted[id=A-2]{We also introduced a {\em modified} FlowNet to detect rip currents (see Appendix A) and compared it with RipViz.}
    
% \end{enumerate}

\deleted[id=A-2]{\vspace{2pt}\noindent\textit{Improving RipViz: } 
We demonstrated that RipViz improved rip current detection compared to previously published methods without requiring any user input at runtime as summarized in Table \ref{table:results_summary}. We further showed that RipViz improved pathline classification for our application domain compared to a state-of-the-art streamline clustering method FlowNet. We optimized the hyper-parameters of our model to sufficiently fit our data. However, for certain hyper-parameters, such as $S$ and $L$, we were not able to conduct broad sensitivity analysis experiments. For instance, for $S$, we could not test values larger than $100$ due to the limitations of our GPU memory. We were also hindered by the time it took to train a single instance of the model ($24-48$ hours). Although we were able to get sufficiently good results by using $S=100$, in the future, we will take a closer look at how the length of the pathline sequence $S$ and training image size $L \times L$ are correlated to the anomaly detection accuracy.}

A key assumption about the stationary videos is that the camera is sufficiently close to the water in order for the optical flow algorithm to pick up measurable velocities.
For similar reasons, we assume that the camera is pointed mostly seaward and not parallel to the beach.
Pointing the camera down a long stretch of beach will create an optical flow that is not representative especially for points farther away from the camera.
Rectifying the frames prior to optical flow calculations may extend the usable range a bit farther but does not justify the extra computational cost.
For these reasons, RipViz works best when the camera is close to the water and pointed mostly seaward.

% We also noticed pathlines from scenes far away from the camera might not move as much as those from a closer location. The pathline movement is limited if the rip current is far from the camera. The pathline may look as if it came from a stationary seed point, similar to one in the sky or the beach. Therefore, we anticipate that RipViz may not be able to detect rip currents in far-a-way scenes. 

Our training data were videos from mostly uncrowded beaches.  
\deleted[id=A1]{If our model is tested against a scene from a crowded beach, pathline sequences from non-water movements e.g. jet ski or beachgoer, the model will likely flag this as anomalous and incorrectly think it's part of a rip current.  We plan to continue making the model more robust via incremental learning as additional data from around the world are collected.}
\added[id=A1]{We did study seed points along the path of a jogger. Such seed points were not marked as anomalous.  This is likely due to the fact that the subsequent pathlines from the sequence marked the initial seed point as mostly normal.  We believe that seed points that are placed on surfers or birds or other objects in the scene will likely be marked as normal as well.  We plan to study this further with more testing data.}

\vspace{2pt}\noindent\textit{Extensions and Future Works: } Some of the contributions listed above are not specific to rip current detection.  We plan to investigate how pathline sequences of unsteady flow fields coupled with the weighted cross-entropy loss function can be used to distinguish between laminar and non-laminar flows, and possibly train a model to detect vortices as anomalous behavior.

We surmise that analysis using pathline sequences may also benefit certain classes
of flow data aside from those in this paper.
In fact, we are exploring that avenue and plan to report on the results in a separate paper.  In short, the particular needs of rip current detection led us to develop the approach presented in this paper, and which also provides another potential tool for the visualization community to study certain classes of flow fields.

% In this paper, we presented RipViz, a deep learning framework for visualizing rip currents. 
% \appendix
% \section{Appendix: Rule Tables for Chapter 2}
% \label{appendix:a}

% \section*{Appendix}
% In \ref{appendix:a}

\section*{Appendix A: Modified FlowNet} \label{appendix:a}
Han et al. \cite{han2018flownet} proposed, FlowNet, a deep learning based method to cluster and select streamlines. They first transformed each streamline into a 1D feature vector of length $1024$. Then they used the t-distributed stochastic neighbor embedding (t-SNE) \cite{tsne} to reduce the dimensionality of the 1D vector
%to a vector of length $2$. 
from $1024$ to $2$.
The resulting vectors are then clustered using DBSCAN\cite{DBSCAN_ester1996density}, a density-based spatial clustering method for applications with noise. The authors provided a user interface to 
%generate the desired clustering 
vary the clustering by changing the maximum distance between two feature descriptors ($eps$) and the minimum number of samples in each cluster ($ms$) in the DBSCAN algorithm.   

We started with the FlowNet architecture for this work, but modified it according to the requirements of our application.
Their primary focus was to cluster streamlines from 3D steady flow fields. Therefore, the authors designed the deep neural network architecture to take in a 3D streamline as input. However, in our application domain, the pathlines are 2D. In order to adapt FlowNet to our application domain, we changed the architecture of FlowNet to take in a 2D pathlines as input. Likewise, we changed the original neural network architecture by replacing 3D convolutional layers and 3D batch normalization layers, respectively, with 2D convolutional layers and 2D batch normalization layers to accommodate 2D pathlines. We also updated the input size and output size of the fully connected layers accordingly. We kept the same number of layers, same activation functions, and the same number of convolutional filters as defined in the FlowNet paper. We refer to the new neural network architecture as \textit{modified} FlowNet.  

We also found that the unweighted binary cross-entropy function used in the FlowNet paper was insufficient to learn the variations of pathlines of the ocean flow for reasons discussed in Section \ref{s:hyper_parameters}. Therefore, we used the weighted binary cross-entropy loss function defined in Section \ref{s:loss_function} when training the \textit{modified} FlowNet. We trained the \textit{modified} FlowNet using the same approach as discussed in the FlowNet paper\cite{han2018flownet}. 

We used the trained \textit{modified} FlowNet to find 1D representations of the pathlines by extracting the latent feature descriptor from the autoencoder. We reduced the dimensionality of the 1D vector by using t-SNE to a vector with a length of $3$. Then we clustered the resulting vectors using DBSCAN, which requires the user to set the maximum distance between two feature descriptors ($eps$) and the minimum number of samples in each cluster ($ms$) before clustering. \added[id=A1]{The cluster with the largest overlap with the ground truth was selected.}\deleted[id=A1]{We colored the seed point of each pathline with a color representing its cluster membership as shown in the second column of Figure \ref{fig:flownet_comparison}. }

\section*{Appendix B: Making the Case for Shorter Pathlines in Noisy Datasets. \label{appendix:b}}
We estimated the flow field from videos using optical flow. We assumed a small error/noise associated with the optical flow estimate. 
Numerical integration, even well designed higher order methods, are subject to accumulation of error.
This problem is aggravated when one is working with noisy datasets.
For the case of pathline integration,
%We also noticed that the longer we integrated the pathline, the higher the error we accumulated in our pathline calculation. Therefore, 
we hypothesized that shorter sequence of pathlines will be more accurate in capturing the flow behavior in noisy flow fields than a single longer pathline.

We used two synthetic datasets to indirectly verify our hypothesis: the \textit{2D unsteady Double Gyre} dataset \cite{Shadden05} and \textit{2D Unsteady Four Rotating Centers} dataset \cite{Guenther17Siggraph}. 

\begin{figure}[tb]
    \centering
     \includegraphics[width=0.99\linewidth]{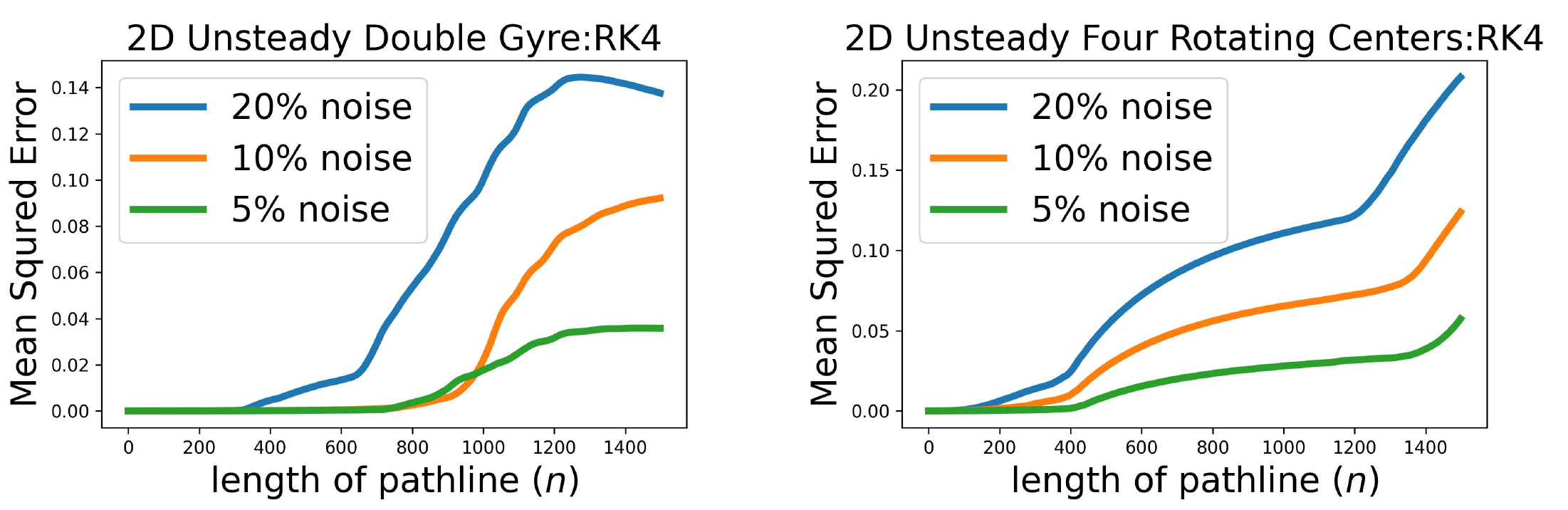}
  \caption{\textbf{Comparison of pathlines with and without noise:} Mean Squared Error (MSE) of pathlines calculated at varying lengths, averaged over 200 randomly seeded pathlines. Notice that for both datasets longer pathlines have higher MSE compared to shorter pathlines.
  %\blue{you sure orange and blue curves are not swapped in 11b?}
  }
  \label{noisy_data_mse} 
\end{figure}

We added noise at $5\%$, $10\%$, and $20\%$ levels to each vector component of each dataset. We compared the similarity of pathlines originating from the same seed point. We observed that shorter pathlines were more similar to those without any noise. To quantify our observations, we randomly seeded $200$ pathlines in each dataset and traced those pathlines. 
%\blue{is it 100 or 200. caption says 200}
We calculated mean squared error (MSE) for pathlines at varying lengths as shown in Figure \ref{noisy_data_mse}. We observed that longer pathlines have higher MSE compared to shorter pathlines. Therefore, we anticipate that shorter pathlines will capture a more accurate representation of the flow behavior in reconstructed noisy flow fields such as ours. 

\begin{figure}[tb]
    \centering
    \includegraphics[width=0.99\linewidth]{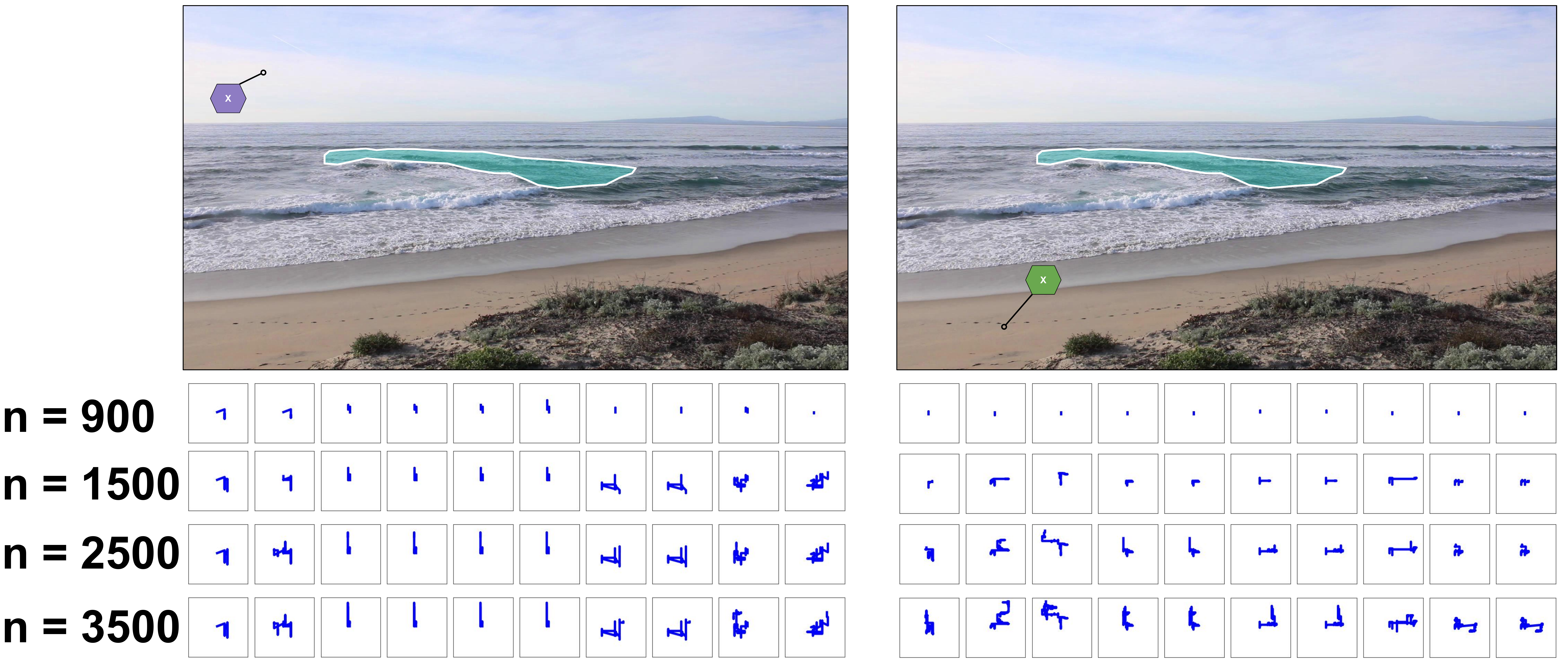}
    \caption{\textbf{Visualization of noise in our dataset:} Here we visualize two pathline sequences, one seeded in the sky (left panel) and the other on the beach (right panel). Each sequence consists of $10$ pathlines. We show the pathlines sequence at four integration time steps(n). We expected these pathlines to be very short and remain closer to the seed point. However, as the pathlines were integrated over a longer period ($n$), they traveled far from the seed point, contradicting our expectation. We attribute this to the accumulation of noise/error when integrating over a longer period in a noisy real world dataset.  }
    \label{fig:noisy_vis}
\end{figure}

\added[id=A-2]{Additionally, we visualized how pathline sequences accumulate errors due to noise over longer integration times as shown in Figure \ref{fig:noisy_vis}. Here, we show two pathline sequences, one seeded in the sky and the other on the beach. We expected these two pathline sequences to be very short and remain closer to the seed point due to the lack of motion around those seed points. However, as we integrate beyond $900$ time steps, we notice that the pathlines seem to travel far from the seed point even though we don't see any visible motion on the video. We attribute this to the accumulation of noise/error when integrating over a more extended time. The noise can come from optical flow estimation or video compression loss.}

% \blue{i think there's also a question of how short. ie. i think you used 100 steps for each pathline.
% if short is good, why not even shorter than 100?}

% \begin{figure}[tb]
%     \centering
%   \subfloat[\label{fig:beads_pathline}]{%
%       \includegraphics[width=0.49\linewidth]{figures/noisy_datasets/beads_example.pdf}}
%     \hfill
%   \subfloat[\label{fig:duffing_pathline}]{%
%         \includegraphics[width=0.49\linewidth]{figures/noisy_datasets/duffing_example_2.pdf}}
%   \caption{Example of pathlines at different levels of noise. All pathlines use the same seedpoint.}
 %   \label{noisy_data_pathlines} 
% \end{figure}

\newpage

%% if specified like this the section will be committed in review mode
\acknowledgments{
This report was prepared in part as a result of work sponsored by
the Southeast Coastal Ocean Observing Regional Association (SECOORA) with NOAA
financial assistance award number NA20NOS0120220. The statements, findings,
conclusions, and recommendations are those of the author(s) and do not necessarily
reflect the views of SECOORA or NOAA.}

\clearpage
\newpage
\bibliographystyle{abbrv-doi-hyperref}

\bibliography{paper}

\end{document}